\documentclass[%
 reprint,
 amsmath,amssymb,
 aps,
 prb,
]{revtex4-1}

\usepackage{graphicx}
\usepackage{dcolumn}
\usepackage{bm}
\usepackage{xcolor}

\begin{document}

\preprint{APS/123-QED}

\title{Rippling of two-dimensional materials by line defects}

\author{Topi K\"ah\"ar\"a}
\author{Pekka Koskinen}%
 \email{pekka.j.koskinen@jyu.fi}
\affiliation{ Nanoscience Center, Department of Physics, 40014 University of Jyv\"askyl\"a}

\date{\today} 

\begin{abstract}
Two-dimensional materials and their mechanical properties are known to be profoundly affected by rippling deformations. However, although ripples are fairly well understood, less is known about their origin and controlled modification. Here, motivated by recent reports of laser-controlled creation of line defects in graphene, we investigate how line defects could be used to control rippling in graphene and other two-dimensional materials. By sequential multi-scale coupling of density-functional tight-binding and continuum elasticity simulations, we quantify the amount of rippling when the number and the cumulative length of the line defects increase. Simulations show that elastic sheets with networks of line defects create rippling that induce considerable out-of-plane ridification and in-plane softening with non-linear elastic behavior. We hope that these insights help to guide experimental attempts to modify the mechanical properties of graphene and other two-dimensional materials. 
\end{abstract}

\maketitle


\section{Introduction}
While graphene and other two-dimensional (2D) materials are often portrayed as planar sheets, real samples frequently contain ripples and other out-of-plane deformations.\cite{meyer_nature_07,lui_nature_09} Rippling deformations affect materials' mechanical, electronic, and thermal properties, and are of pivotal importance for a number of applications.

From a mechanical applications point of view, the most notable effect of rippling is its propensity to increase the out-of-plane rigidity of 2D materials.\cite{Nicholl2015,AKINWANDE201742,Lee385,doi:10.1021/nl301080v} The magnitude of rigidity is important for kirigami applications~\cite{Blees2015}, nanomechanical systems~\cite{zalalutdinov_NL_12}, mechanical resonators~\cite{Bunch2007}, cantilevers~\cite{Nicholl2017}, material systems~\cite{tapaszto_nphys_12}, and applications dependent on the thermal expansion coefficient~\cite{chen_PRB_11}, to mention six examples. Rigidity is equally importance also for nanostructured membranes, nanoribbons in particular.\cite{neek-amal_PRB_10} The application potential of customized rigidity is aptly illustrated by the mundane example of corrugated sheets as packaging materials; customized rigidity of 2D materials can be envisioned to substantially expand the range of mechanical applications at the nanoscale.\cite{Koskinen2014} 

In addition to mechanical properties, rippling affects also phononic and electronic transport. Directional rippling renders phononic transport anisotropic~\cite{Wang2014b}, while electronic transport is modified by pseudo-magnetic fields induced by areas of local Gaussian curvature.\cite{Katsnelson2008,kim_EPL_08,castro_neto_RMP_09,DasSarma2011} 

Finally, rippling changes the nature of interaction with substrates. Rigidified membranes conform poorly to the surface morphology and come into local, dispersed contacts with the substrate, reducing adhesion.\cite{Wang2015a} Rippling can therefore be used to customize surface adhesion, adjust the extent of intercalation of foreign molecules, and modify various properties of multilayers.\cite{He2013a,Gao2015a} 

Ripples can have several different origins. They can originate from thermal fluctuations, defects, adsorbates, and external stresses and their characteristics have been investigated by atomic force and transmission electron microscopies.\cite{Wang2012b,Deng2016a,AKINWANDE201742,thompson-flagg_EPL_09,GAO201442,kholmanov_PRB_09,PhysRevB.93.125431,doi:10.1021/nn102598m,Lee1073,Bao2009,Liu2011,Duan2011,Ludacka2018} But whatever the origin, the list of applications above implies that achieving better control over rippling should be considered highly desirable. 

It is therefore exciting that recent experiments have shown indications of controlled rigidification of graphene by using a technique called optical forging.\cite{Johansson2017, Koskinen2018a}. The technique, which consists of direct writing with pulsed laser under inert atmosphere, creates defect structures of linear character.\cite{Koivistoinen2016,Hiltunen2020} The graphene samples in these experiments were initially supported, but in the resulting three-dimensional structures the forged graphene is essentially suspended, making the presence of substrates irrelevant. Modeling and experiments indicate that the forging process presumably creates line defects as arrays of adjacent Stone-Wales point defects.\cite{Hiltunen2020} Although the mechanical properties of rippled membranes as such are fairly well known~\cite{PhysRevE.88.012136}, line defects as the source of rippling is not understood.

In this article, therefore, we pursue to investigate how line defects create rippling in 2D materials, by using a sequential multiscale approach. We start from atomistic modeling, using Stone-Wales line defects in graphene as prototypical, exemplary line defects. Results from these atomistic simulations are then fed into a mesoscale continuum model that enables investigating rippling at relevant length scales and makes the results generic to different 2D materials.

\section{Elastic parameters from atomistic simulations}
\label{atomistic_sim}

\begin{figure}[t!]
    \centering
    \includegraphics[width=0.9\columnwidth]{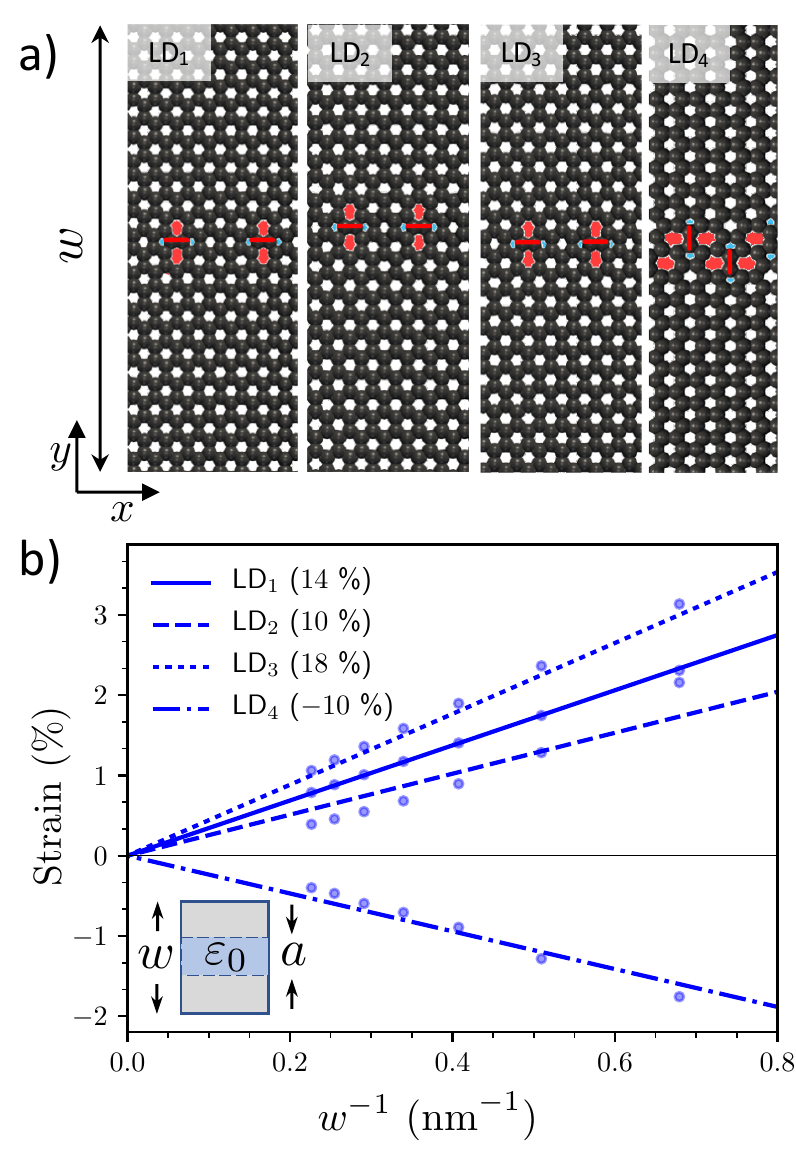}
    \caption{Mesoscopic elasticity of line defects from atomistic modeling. a) Four Stone-Wales line defect (LD) models, enclosed in simulation cells of width $w$ from $1.6$ to $4.7$~nm. Cell length $l_x$ was optimized for each $w$ to get the strain $\varepsilon(w)=\Delta l_x/l_x$ along the line defect. The rotated bonds (red bars), surrounded by pentagons (blue shading) and heptagons (red shading), indicate the direction of compressive stress. Line defect models are all planar except for LD$_2$, where the peak-to-peak out-of-plane corrugation is $0.9$~\AA. b) Unit cell strains $\varepsilon(w)$ for each line defect model. Strains are fitted to the functional form $\varepsilon(w)=\varepsilon_0(a/w)$, where $\varepsilon_0$ (in brackets) is a line defect -dependent fit parameter and $a=2.46$~\AA\ is the graphene lattice constant. Inset: the effect of the line defect on the mesoscale is equivalent to a strip of width $a$ strained longitudinally by $\varepsilon_0$.}
    \label{fig:dftb}
\end{figure}

To obtain parameters for the continuum model, we first investigate atomistic models for prototypical line defects. We consider line defects induced into pristine graphene, excluding grain boundaries and other topological defects that change crystal orientation.\cite{Liu2010a,malola_PRB_10} As discussed above, we focus on line defects inspired by observations from optically forged graphene samples that presumably contain linear arrays of Stone-Wales (SW) defects.\cite{Ma2009,Hiltunen2020} Although these line defects are neither fully characterized nor fully established, they enable the construction of concrete and feasible line defect models. In SW defects one carbon bond rotates $90$ degrees and forms two neighboring pentagons and heptagons. The formation energy is large ($4.6$~eV) \cite{Ma2009}, but it reduces by as much as $1.5$~eV when a second defect is formed near an existing one at suitable distance and orientation.\cite{Fan2010,Hiltunen2020} This attractive and anisotropic interaction makes pulsed laser irradiation auspicious for growing SW line defects.\cite{Hiltunen2020} Since the atomic arrangement of SW line defects is unknown, we content ourselves for creating four different model geometries and obtain rough magnitudes for the microscopic strain fields involved (Fig.~\ref{fig:dftb}a). 

The model geometries are simulated by density-functional tight-binding (DFTB) theory \cite{porezag_PRB_95,elstner_PRB_98,frauenheim_PSSb_00}, using the \texttt{hotbit} code.\cite{koskinen_CMS_09}. DFTB was chosen because it reproduces graphene's elastic properties well compared to density-functional theory (bending modulus $1.6$~eV and Young's modulus $1.4$~TPa)~\cite{Memarian2015,kudin_PRB_01,koskinen_PRB_10b} and enables effective explorations across various system sizes by a few orders of magnitude faster calculations.\cite{Frauenheim2002,koskinen_PRB_10,koskinen_PRL_10,koskinen_PRB_12,kit_PRB_12,Ramasubramaniam2012,koskinen_APL_11,kit_PRB_11,Korhonen2014a,Koskinen2016}

The line defects were modeled in a periodic, rectangular simulation cell of width $w$ and length $l_x$ (Fig.~\ref{fig:dftb}a), within which the SW defects were distributed along the $x$-direction. Using a $10\times 1$ $k$-point sampling, the geometries were optimized to force tolerance $0.005$~eV/\AA\ \cite{bitzek_PRL_06} and zero unit cell stress. The ultimate result from this procedure was the unit cell strain $\varepsilon(w)=\Delta l_x/l_x$ for different widths $w$. 

As the central result, the unit cell strains were observed to scale as 
\begin{equation}
\varepsilon(w)=\varepsilon_0\frac{a}{w},
\label{eq:strain}
\end{equation}
where $a=2.46$~\AA\ is the graphene lattice constant and $\varepsilon_0$ is a fit parameter characterizing the line defect in question (Fig.~\ref{fig:dftb}b). Conceptually, the functional form implies that---from a continuum point of view---the presence of an atomistic line defect can be simply and accurately represented by a stripe of width $a$ with a longitudinal strain $\varepsilon_0$ (inset of Fig.~\ref{fig:dftb}b). The strain $\varepsilon_0$ for different line defect models varied between $\varepsilon\approx -10\ldots 20$~\%. 
While $\varepsilon_0$ has both negative and positive values, here we focus on positive values, meaning line defects with compressive stress. Only compressive stress can be released by rippling deformations when the defect density is low. Such rippling is demonstrated at atomic scale by the planar model LD$_3$, in which the compressive stress is reduced by a slight out-of-plane buckling (Fig.~\ref{fig:dftb}b). Conversely, tensile stress can be released by rippling only at defect densities high enough to fall beyond the scope of this work. One example of a line defect with tensile stress is LD$_4$, in which rotated bonds and the compressive stress orient perpendicular to the line defect.\cite{Koskinen2008,Koskinen2009a} 

To summarize the atomistic simulations, the central results are the scaling in Eq.~(\ref{eq:strain}) and the range of strains $\varepsilon_0\sim 10-20$~\%. There would have been many more possibilities for atomic structures of line defects, but obtaining a rough range of strains suffices for our purposes. Knowledge of the range allows us to proceed to mesoscale continuum elasticity modeling, which can use Eq.~(\ref{eq:strain}) to account for the presence of line defects consistently with respect to the discretization length. 

\section{Continuum elasticity simulations}

\subsection{The elasticity model}

The continuum elasticity model we use is similar to the one introduced in Ref.~\onlinecite{PhysRevA.38.1005} and used in Ref.~\onlinecite{kit_PRB_12} in the context of carbon nanotubes. The sheet is modeled as a hexagonal lattice of linear springs of equilibrium length $d$ and spring constant $\tilde{k}_s$. The in-plane strain energy of a spring connecting vertices A and B is thus
\begin{equation}
    E_s = \frac{1}{2}\tilde{k}_s(|\vec{r}_{AB}|-d)^2,
\end{equation}
where $|\vec{r}_{AB}|$ is the distance from A to B. The bending energy related to neighboring vertices A and B is given by 
\begin{equation}
    E_b = \frac{1}{2}\tilde{k}_b\theta_A^2 + \frac{1}{2}\tilde{k}_b\theta_B^2,
\end{equation}
where $\tilde{k}_b$ is a parameter controlling the bending rigidity and $\theta_X$ is the angle between the vector $\vec{r}_{AB}$ and the unit normal vector $\hat{n}_X$ at vertex $X\in \{A,B\}$. The normal vector $\hat{n}_X$ is defined as the area-weighted mean of the normal vectors of the triangles surrounding the vertex.

In the continuum limit of the model, the energy density for small uniaxial strains $\varepsilon$ is approximately 
\begin{equation}
    F_s = \frac{3\sqrt{3}}{8}\tilde{k}_s\varepsilon^2 \equiv \frac{1}{2}k_s\varepsilon^2.
    \label{eq:Fs}
\end{equation}
Similarly, the energy density for pure bending at bending radius $R$ is
\begin{equation}
    F_b = \frac{3\sqrt{3}}{16}\frac{\tilde{k}_b}{R^2} \equiv \frac{1}{2}\frac{k_b}{R^2}.
    \label{eq:Fb}
\end{equation}
The quantities $k_s$ and $k_b$ in Eqs.~(\ref{eq:Fs}) and (\ref{eq:Fb}) are the uniaxial strain modulus and the bending modulus of the sheet, to be compared with literature values.

This model was implemented in Python using a rhombic cell with periodic boundary conditions. The cell length was $l_\text{cell}=120$~nm and discretization length $2$~nm (unstretched spring lengths). These values are sufficient for a faithful continuum description of the type of rippling we focus on. The strain modulus was $k_s=26$~eV/Å$^2$, which is comparable to the strain modulus of graphene while represents roughly also other 2D materials such as hexagonal boron nitride and transition metal dichalcogenides.\cite{AKINWANDE201742} In order to obtain different ratios for $k_s/k_b$ and thereby represent materials with various rigidities~\cite{AKINWANDE201742,koskinen_PRB_10b,Koskinen2014a,Yu2016}, the bending modulus $k_b$ was chosen the values of $13$, $26$, and $52$~eV. The bending moduli are greater than inferred from atomistic \emph{ab initio} calculations for flat graphene, but still significantly smaller than values measured for graphene samples and expected to be realistic considering the discretization length.\cite{AKINWANDE201742,doi:10.1021/nl301080v} Reiterating, although our atomistic simulations involved only graphene and certain line defect models, we use generic parameters with the intention to get a broad view of the effect of line defects on the rippling on 2D materials; our elasticity model is fully general.

In the model, the strain fields from line defects were represented by the stretching of bonds. The magnitude of stretching was determined by adopting a sequential multi-scale strategy, which implied that parallel strains induced by atomic scale defects are inversely proportional to the width of the system, as given by Eq.~(\ref{eq:strain}). Then, for a given distribution of line defects and the concomitant strain field, the minimum energy geometry was determined by the L-BFGS-B optimization method with a relative accuracy of $10^{-7}$, meaning total energy tolerance below $0.1$~eV. Also the cell size was optimized for each strain field. Using the Nelder-Mead simplex method implemented in the SciPy optimization library \cite{2020SciPy-NMeth}, cell size was optimized to $0.1$~\AA\ accuracy.

The effective elastic moduli of the rippled sheets---the actual measurable quantities in typical experiments---were calculated by modifying cell sizes and boundary conditions. The effective strain moduli $k_s^\text{eff}$ were calculated by straining the cells uniaxially to $4$~\%\ maximum strain. The effective bending moduli $k_b^\text{eff}$ were calculated by adapting the boundary conditions to cylindrical symmetry, in the spirit of revised periodic boundary conditions.\cite{koskinen_PRL_10,kit_PRB_11} The smallest radii for determining the bending modulus were $10$ times the cell length. For these small strains and small curvatures the effective moduli could then be extracted directly from the energy density profiles through Eqs.~(\ref{eq:Fs}) and (\ref{eq:Fb}).

To get progressively deepening insight into the effect of line defects on rippling, we investigated three different line defect scenarios: \emph{i}) an isolated line defect, \emph{ii}) a hexagonal line defect lattice, and \emph{iii}) a random network of line defects. We investigated these scenarios using different values of compressive strain $\varepsilon_0$ and increasing cumulative lengths of the line defects.

\subsection{Limit of low defect density: isolated line defects}

\begin{figure}[tb!]
    \centering
    \includegraphics[width=0.9\columnwidth]{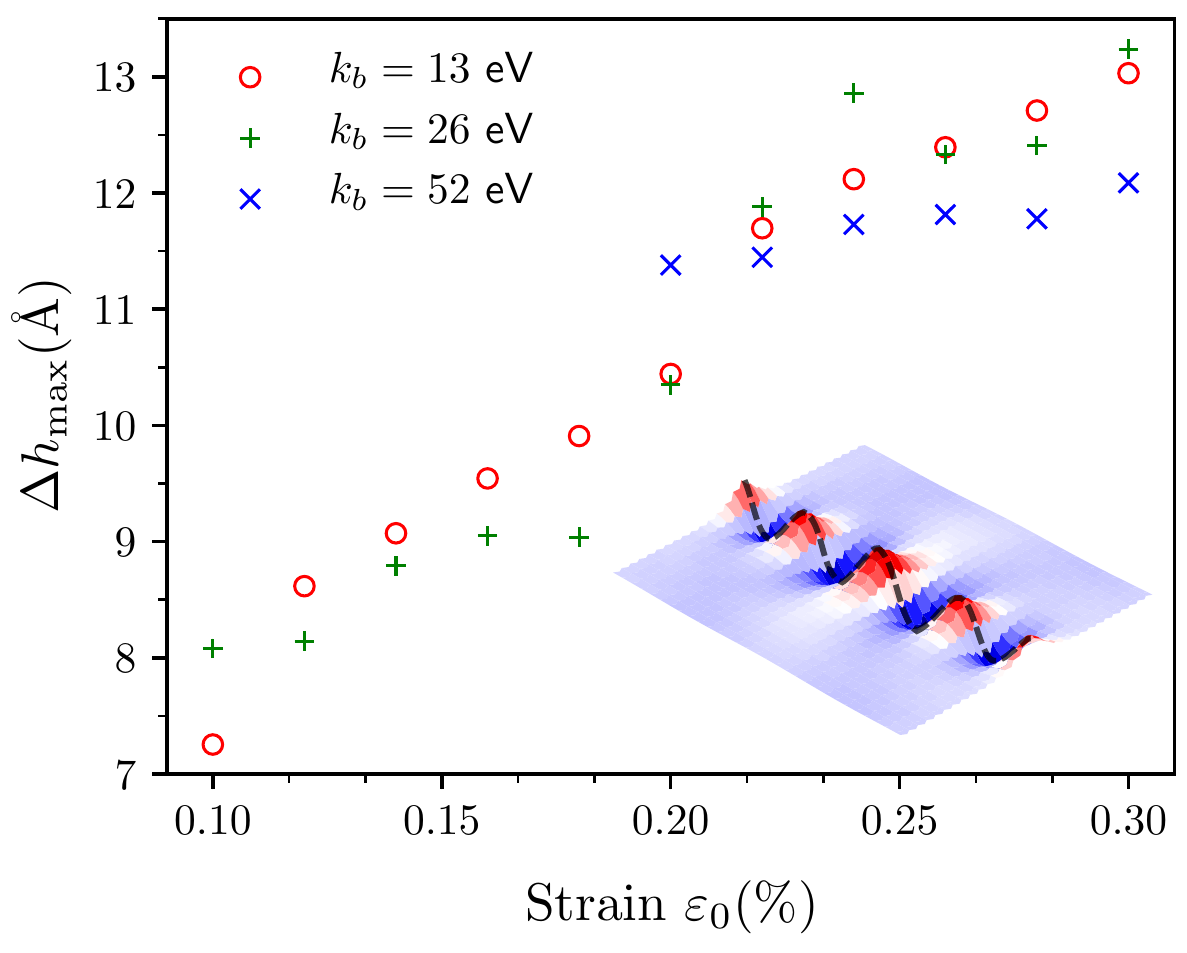}
    \caption{Rippling with an isolated line defect. Plot shows the maximum height variation of the sheet $\Delta h_\text{max}$ as a function of $\varepsilon_0$ for three different values of $k_b$. Inset: Optimized geometry of an isolated line defect (dashed line) with $\varepsilon_0=10$~\%\ and $k_b=26$~eV.}
    \label{fig:2}
\end{figure}

In the first scenario we had a single, isolated defect spanning across the sheet. We chose the width $w$ in Eq.~(\ref{eq:strain}) equal to the lattice constant of $2$~nm, representing a just transformation of given $\varepsilon_0$ into appropriate mesoscopic strain. An illustrative sample of the resulting optimized structure reveals rippling surrounding the line defect (Fig.~\ref{fig:2}). This type of rippling is familiar from the everyday behavior of clothes and fabrics. It also agrees with the reported rippling at the edges of semi-infinite graphene membranes.\cite{Shenoy2008,Shenoy2010} The ripple amplitude increases as a function of $\varepsilon_0$. The increase is steady, although slightly non-monotonous due to finite size effects; the cell can accommodate only a discrete number of waves along the line defect. Note that, at least within the given strain range, the bending modulus has only a small effect on the ripple amplitude. This trivial behavior indicates that the geometry is dominated by strain energy. 

\subsection{Interacting defects: hexagonal and random line defect networks}

\begin{figure}[tb!]
    \centering
    \includegraphics[width=0.9\columnwidth]{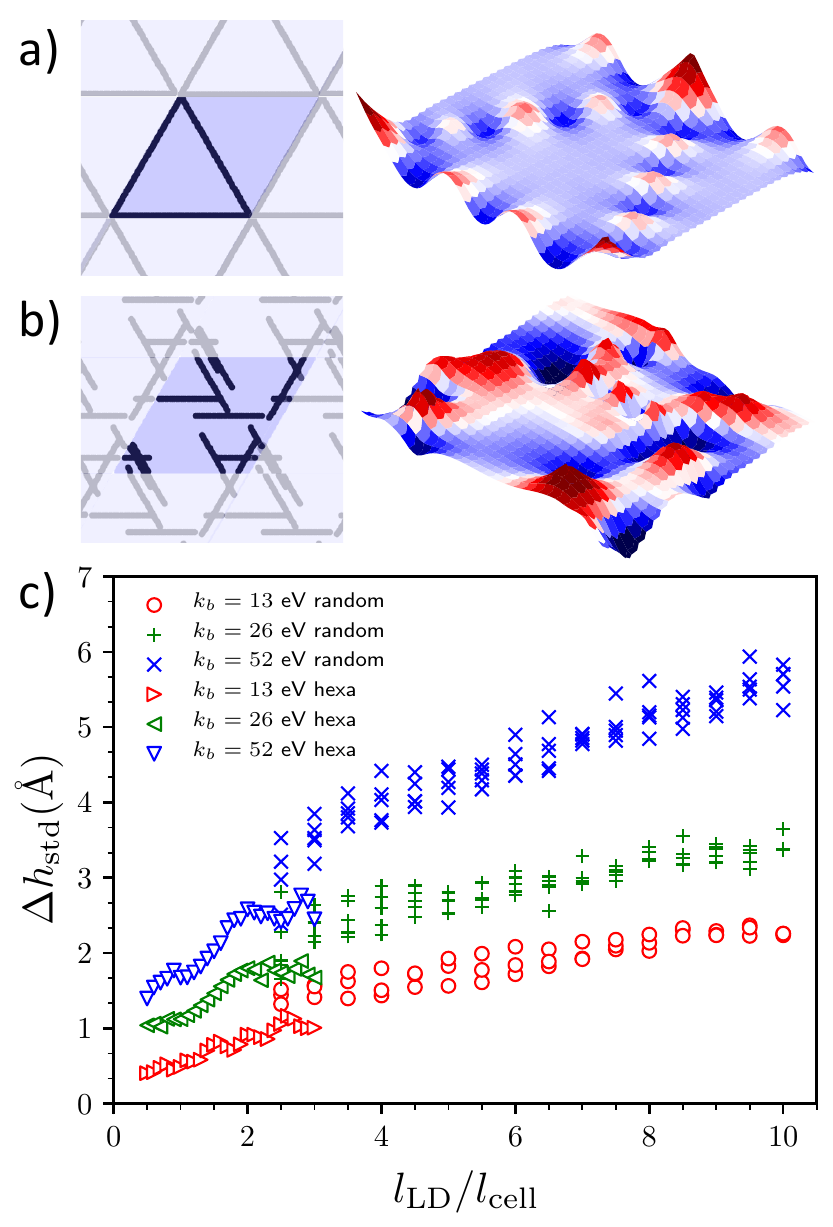}
    \caption{Rippling with hexagonal and random line defect networks. a) The geometry of a hexagonal line defect network with cumulative length of line defects $l_\text{LD}=3\times l_\text{cell}$ (left), together with the resulting rippling for $k_b=26$~eV (right). The copies of the simulation cell on the left are shaded. b) The geometry of a random line defect network with $l_\text{LD}=5\times l_\text{cell}$ (left), together with the resulting ripples for $k_b=26$~eV (right). c) Standard deviation of sheet height as a function of the cumulative length of line defects for different scenarios of networks and values of $k_b$.}
    \label{fig:3}
\end{figure}

Hexagonal and random line defect network scenarios yield similar results and we therefore discuss them simultaneously. Here the main variable is $l_\text{LD}$, the cumulative length of all the line defects, and especially its ratio to the cell length $l_\text{cell}$. In the hexagonal scenario, defects were grown from single points toward six symmetric directions, increasing the defect length until the lines became fully connected, with maximum cumulative length of $l_\text{LD}=3\times \l_\text{cell}$ (Fig.~\ref{fig:3}a). In the random line defect scenario the cumulative line defect length could be larger and was varied from $2.5$ to $10\times l_\text{cell}$. Random line defect networks were investigated with $10$, $20$, and $30$ separate line defects with both uniform and linear length distributions, in order to mimic the type of networks proposed in Ref.~\onlinecite{Hiltunen2020} (Fig.~\ref{fig:3}b). It turned out that neither the number of line defects nor the type of their length distribution made noticeable difference in the trends; below we show results for all defect numbers and distributions.

As in the previous subsection, both hexagonal and random scenarios are simulated for three values of bending moduli. We used $\varepsilon_0=10$~\%\ with $k_b=13$ and $26$~eV and $\varepsilon_0=20$~\%\ with $k_b=52$~eV. Larger $\varepsilon_0$ for the largest $k_b$ was necessary because a stiff sheet made it difficult to discern reliable trends in those situations when a small $\varepsilon_0$ was supplemented by small $l_\text{LD}$.

As a clear trend for both scenarios, the rippling intensifies steadily upon increasing cumulative length of line defects. The intensifying is evident in the increase of the standard deviation in the height distribution (Fig.~\ref{fig:3}c). Standard deviation is a robust quantity easy to obtain from the model and determinable also from experiments, unlike peak-to-peak amplitudes that are subject to wild fluctuations. Especially the hexagonal scenario shows peaked corners, while other areas remain only modestly rippled (Fig.~\ref{fig:3}a). For both scenarios the standard height deviation increases roughly linearly with increasing $l_\text{LD}$, although for the hexagonal scenario it is noticeably smaller, visible around $l_\text{LD}/l_\text{cell}\approx 3$ where the data from the two scenarios overlap. This difference arises presumably because in the hexagonal scenario the line defects with given $l_\text{LD}$ have maximal spatial separation and are consequently less prone to spread ripples effectively throughout the sheet.

When $l_\text{LD}$ increases, also the total surface area of the sheet expands. This area is sometimes referred to as the hidden area, as it does not necessarily result it actual lateral expansion.\cite{Nicholl2017} Indeed, as a consequence of the very low bending stiffness compared to strain modulus, the sheet prefers rippling over lateral expansion. Similar phenomenology is responsible for the negative thermal expansion coefficient of graphene.\cite{balandin_nmat_11} However, with the given material parameters the changes in lateral cell dimensions were small enough to render the nature of lateral behavior inconclusive.


Finally, we proceed to the main results, which discuss how line defect networks---through rippling---affect 2D materials' effective elastic properties.

\begin{figure}[tb!]
    \centering
    \includegraphics[width=0.9\columnwidth]{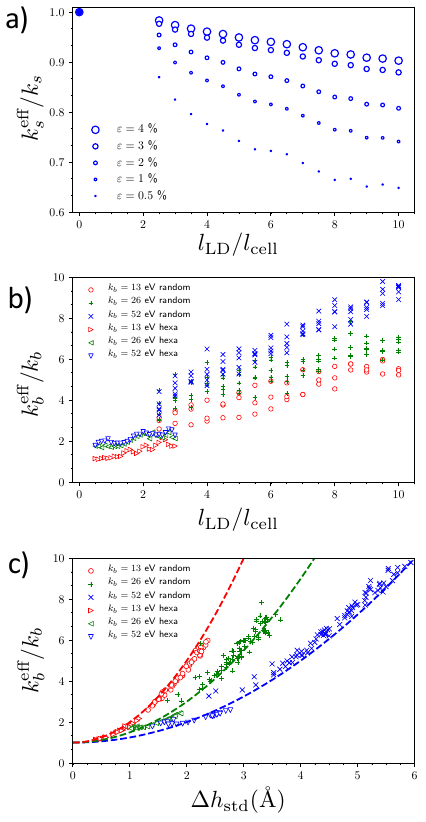}
    \caption{Effective elastic moduli of elastic sheets with line defect networks. a) Relative changes in effective strain moduli ($k_s^\text{eff}/k_s$) of rippled sheets as a function of the cumulative line defect length $l_\text{LD}$ for random line defect networks, demonstrating non-linear elasticity. The effect of $k_b$ on $k_s^\text{eff}$ was small, whereby the values were averaged over different $k_b$. b) Relative changes in effective bending moduli ($k_b
   ^\text{eff}/k_b$) of rippled sheets as a function of the cumulative line defect length, for different $k_b$ and scenarios of line defect networks. c) The relative change in the effective bending modulus ($k_b^\text{eff}/k_b)$ as a function of rippling intensity measured by the standard height deviation. The dashed curves are given by Eq.~(\ref{eq:kbr}).}
    \label{fig:4}
\end{figure}

Among the most notable results is the highly non-linear elastic behavior of strained sheets. This non-linearity can be seen as a strongly strain-dependent effective strain modulus $k_s^\text{eff}$ (Fig.~\ref{fig:4}a).\cite{Nicholl2017} The modulus at strains around $0.5$~\%\ is tens of percents smaller than at strains around $4$~\%. This trend can be understood in geometric terms: small strains involve the flattening of the initial ripples and concerns mostly (cheap) bending energy, while large strains involve the further stretching of the already flattened sheet and concerns mostly (expensive) stretching energy. Thus, at the limit of large strain, the effective strain modulus necessarily approaches $k_s$. This geometrical picture is consistent with the decrease of $k_s^\text{eff}/k_s$ upon increasing $l_\text{LD}$ (compare Figs.~\ref{fig:3}c and \ref{fig:4}a).

In contrast to the non-linear behavior of the effective strain modulus, the effective bending modulus turned out to be a well-defined quantity, independent of the applied bending radius. The effective bending modulus shows a clear increasing trend upon increasing $l_\text{LD}$ (Fig.~\ref{fig:4}b). This trend is much expected, in view of the well-known stiffening effect of ripplings in thin membranes.\cite{PhysRevE.88.012136} Interestingly, the slope between $k_b^\text{eff}/k_b$ and $l_\text{LD}/l_\text{cell}$ is roughly one, although for larger $k_b$ the stiffening is seen more pronounced. Moreover, around $l_\text{LD}/l_\text{cell}\approx 3$, the effective bending modulus is larger with random line defects than with hexagonal line defects, in line with a corroborating trend in standard height deviation (Fig.~\ref{fig:3}c).

The origin for the trend in the effective bending modulus is illustrated particularly well when $k_b^\text{eff}$ is plotted as a function of the standard deviation $\Delta h_\text{std}$ of the sheet height (Fig.~\ref{fig:4}c). The relation between $k_b^\text{eff}/k_b$ and $\Delta h_\text{std}$ fits well with the quadratic expression
\begin{equation}
\frac{k_b^\text{eff}}{k_b} = 1 + \frac{1}{2}\frac{k_s}{k_b}\Delta h_{\textrm{std}}^2.
\label{eq:kbr}
\end{equation}
The expression can be justified by a simple dimensional analysis and the limit $k_b^\text{eff}\rightarrow k_b$ as $\Delta h_\text{std}\rightarrow 0$; the factor $\tfrac{1}{2}$ is a (convenient) fit parameter. This expression is consistent with the small ripple amplitude result from Ref.~\onlinecite{PhysRevE.88.012136}, but inconsistent with the asymptotic scaling for large ripples predicted in the same paper. We therefore expect the relation (\ref{eq:kbr}) to become invalid at larger scales. Nevertheless, the spatial character of rippling caused by line defects, even if randomly displaced, may differ from the spatial character underlying random rippling assumed in Ref.~\onlinecite{PhysRevE.88.012136}. The relation (\ref{eq:kbr}), the spatial character of rippling, and the nature of rigidification thus deserve further investigations.

\section{Summary and conclusions}

In this work we used sequential multi-scale modeling to investigate the influence of line defects on the mesoscopic elastic properties of two-dimensional materials. Inspired by recent experiments, we used arrays of Stone-Wales defects as the prototypical models of non-topological, atomic scale line defects.\cite{Hiltunen2020} Density-functional tight-binding calculations were used to obtain reasonable parameters for mesoscale continuum modeling by using the scaling law of Eq.~(\ref{eq:strain}). The mesoscale modeling was then framed at couple of orders of magnitude larger length scales and with parameters that reflected our attempt to obtain generic results for the elastic properties representing various 2D materials.

The modeling showed that sheets ripple predictably as the number, the intrinsic strain $\varepsilon_0$, and the cumulative length of line defects $l_\text{LD}$ increase (Figs.~\ref{fig:2} and \ref{fig:3}). The increased rippling consequently reduces the effective elastic modulus and substantially increase the effective bending modulus of the sheet (Fig.~\ref{fig:4}), in line with findings in earlier reports. These results are well in line with the experimentally observed disparity between atomic and mesoscale elastic properties of 2D materials.\cite{Blees2015} We also provided a useful relationship (\ref{eq:kbr}) between the rigidification of the sheet and the magnitude of its ripples due to the presence of line defects. 

In conclusion, the multiscale model has turned out useful in bridging the gap between the atomic and mesoscale elastic properties of 2D materials. While the atomic models were specific to graphene, the results were made generic for various atomically thin materials with comparable parameters. Should experiments discover mechanisms to create line defects in materials other than graphene, the multiscale formalism could readily be used to explore their quantitative influence on mesoscale mechanical properties. However, we hope5 that already the results above will give insight into the influence of rippling by line defects and thereby further guide experimental attempts to modify the mechanical properties of 2D materials. 

\acknowledgements{}

We acknowledge the Academy of Finland for funding (project 297115).


\begin{thebibliography}{70}%
	\makeatletter
	\providecommand \@ifxundefined [1]{%
		\@ifx{#1\undefined}
	}%
	\providecommand \@ifnum [1]{%
		\ifnum #1\expandafter \@firstoftwo
		\else \expandafter \@secondoftwo
		\fi
	}%
	\providecommand \@ifx [1]{%
		\ifx #1\expandafter \@firstoftwo
		\else \expandafter \@secondoftwo
		\fi
	}%
	\providecommand \natexlab [1]{#1}%
	\providecommand \enquote  [1]{``#1''}%
	\providecommand \bibnamefont  [1]{#1}%
	\providecommand \bibfnamefont [1]{#1}%
	\providecommand \citenamefont [1]{#1}%
	\providecommand \href@noop [0]{\@secondoftwo}%
	\providecommand \href [0]{\begingroup \@sanitize@url \@href}%
	\providecommand \@href[1]{\@@startlink{#1}\@@href}%
	\providecommand \@@href[1]{\endgroup#1\@@endlink}%
	\providecommand \@sanitize@url [0]{\catcode `\\12\catcode `\$12\catcode
		`\&12\catcode `\#12\catcode `\^12\catcode `\_12\catcode `\%12\relax}%
	\providecommand \@@startlink[1]{}%
	\providecommand \@@endlink[0]{}%
	\providecommand \url  [0]{\begingroup\@sanitize@url \@url }%
	\providecommand \@url [1]{\endgroup\@href {#1}{\urlprefix }}%
	\providecommand \urlprefix  [0]{URL }%
	\providecommand \Eprint [0]{\href }%
	\providecommand \doibase [0]{https://doi.org/}%
	\providecommand \selectlanguage [0]{\@gobble}%
	\providecommand \bibinfo  [0]{\@secondoftwo}%
	\providecommand \bibfield  [0]{\@secondoftwo}%
	\providecommand \translation [1]{[#1]}%
	\providecommand \BibitemOpen [0]{}%
	\providecommand \bibitemStop [0]{}%
	\providecommand \bibitemNoStop [0]{.\EOS\space}%
	\providecommand \EOS [0]{\spacefactor3000\relax}%
	\providecommand \BibitemShut  [1]{\csname bibitem#1\endcsname}%
	\let\auto@bib@innerbib\@empty
	\bibitem [{\citenamefont {Meyer}\ \emph {et~al.}(2007)\citenamefont {Meyer},
		\citenamefont {Geim}, \citenamefont {Katsnelson}, \citenamefont {Novoselov},
		\citenamefont {Booth},\ and\ \citenamefont {Roth}}]{meyer_nature_07}%
	\BibitemOpen
	\bibfield  {author} {\bibinfo {author} {\bibfnamefont {J.~C.}\ \bibnamefont
			{Meyer}}, \bibinfo {author} {\bibfnamefont {A.~K.}\ \bibnamefont {Geim}},
		\bibinfo {author} {\bibfnamefont {M.~I.}\ \bibnamefont {Katsnelson}},
		\bibinfo {author} {\bibfnamefont {K.~S.}\ \bibnamefont {Novoselov}}, \bibinfo
		{author} {\bibfnamefont {T.~J.}\ \bibnamefont {Booth}},\ and\ \bibinfo
		{author} {\bibfnamefont {S.}~\bibnamefont {Roth}},\ }\bibfield  {title}
	{\bibinfo {title} {{The structure of suspended graphene sheets}},\
	}\href@noop {} {\bibfield  {journal} {\bibinfo  {journal} {Nature}\ }\textbf
		{\bibinfo {volume} {446}},\ \bibinfo {pages} {60} (\bibinfo {year}
		{2007})}\BibitemShut {NoStop}%
	\bibitem [{\citenamefont {Lui}\ \emph {et~al.}(2009)\citenamefont {Lui},
		\citenamefont {Liu}, \citenamefont {Mak}, \citenamefont {Flynn},\ and\
		\citenamefont {Heinz}}]{lui_nature_09}%
	\BibitemOpen
	\bibfield  {author} {\bibinfo {author} {\bibfnamefont {C.~H.}\ \bibnamefont
			{Lui}}, \bibinfo {author} {\bibfnamefont {L.}~\bibnamefont {Liu}}, \bibinfo
		{author} {\bibfnamefont {K.~F.}\ \bibnamefont {Mak}}, \bibinfo {author}
		{\bibfnamefont {G.~W.}\ \bibnamefont {Flynn}},\ and\ \bibinfo {author}
		{\bibfnamefont {T.~F.}\ \bibnamefont {Heinz}},\ }\bibfield  {title} {\bibinfo
		{title} {{Ultraflat graphene}},\ }\href@noop {} {\bibfield  {journal}
		{\bibinfo  {journal} {Nature}\ }\textbf {\bibinfo {volume} {462}},\ \bibinfo
		{pages} {339} (\bibinfo {year} {2009})}\BibitemShut {NoStop}%
	\bibitem [{\citenamefont {Nicholl}\ \emph {et~al.}(2015)\citenamefont
		{Nicholl}, \citenamefont {Conley}, \citenamefont {Lavrik}, \citenamefont
		{Vlassiouk}, \citenamefont {Puzyrev}, \citenamefont {Sreenivas},
		\citenamefont {Pantelides},\ and\ \citenamefont {Bolotin}}]{Nicholl2015}%
	\BibitemOpen
	\bibfield  {author} {\bibinfo {author} {\bibfnamefont {R.~J.}\ \bibnamefont
			{Nicholl}}, \bibinfo {author} {\bibfnamefont {H.~J.}\ \bibnamefont {Conley}},
		\bibinfo {author} {\bibfnamefont {N.~V.}\ \bibnamefont {Lavrik}}, \bibinfo
		{author} {\bibfnamefont {I.}~\bibnamefont {Vlassiouk}}, \bibinfo {author}
		{\bibfnamefont {Y.~S.}\ \bibnamefont {Puzyrev}}, \bibinfo {author}
		{\bibfnamefont {V.~P.}\ \bibnamefont {Sreenivas}}, \bibinfo {author}
		{\bibfnamefont {S.~T.}\ \bibnamefont {Pantelides}},\ and\ \bibinfo {author}
		{\bibfnamefont {K.~I.}\ \bibnamefont {Bolotin}},\ }\bibfield  {title}
	{\bibinfo {title} {{The effect of intrinsic crumpling on the mechanics of
				free-standing graphene}},\ }\href {https://doi.org/10.1038/ncomms9789}
	{\bibfield  {journal} {\bibinfo  {journal} {Nature Communications}\ }\textbf
		{\bibinfo {volume} {6}},\ \bibinfo {pages} {1} (\bibinfo {year}
		{2015})}\BibitemShut {NoStop}%
	\bibitem [{\citenamefont {Akinwande}\ \emph {et~al.}(2017)\citenamefont
		{Akinwande}, \citenamefont {Brennan}, \citenamefont {Bunch}, \citenamefont
		{Egberts}, \citenamefont {Felts}, \citenamefont {Gao}, \citenamefont {Huang},
		\citenamefont {Kim}, \citenamefont {Li}, \citenamefont {Li}, \citenamefont
		{Liechti}, \citenamefont {Lu}, \citenamefont {Park}, \citenamefont {Reed},
		\citenamefont {Wang}, \citenamefont {Yakobson}, \citenamefont {Zhang},
		\citenamefont {Zhang}, \citenamefont {Zhou},\ and\ \citenamefont
		{Zhu}}]{AKINWANDE201742}%
	\BibitemOpen
	\bibfield  {author} {\bibinfo {author} {\bibfnamefont {D.}~\bibnamefont
			{Akinwande}}, \bibinfo {author} {\bibfnamefont {C.~J.}\ \bibnamefont
			{Brennan}}, \bibinfo {author} {\bibfnamefont {J.~S.}\ \bibnamefont {Bunch}},
		\bibinfo {author} {\bibfnamefont {P.}~\bibnamefont {Egberts}}, \bibinfo
		{author} {\bibfnamefont {J.~R.}\ \bibnamefont {Felts}}, \bibinfo {author}
		{\bibfnamefont {H.}~\bibnamefont {Gao}}, \bibinfo {author} {\bibfnamefont
			{R.}~\bibnamefont {Huang}}, \bibinfo {author} {\bibfnamefont {J.-S.}\
			\bibnamefont {Kim}}, \bibinfo {author} {\bibfnamefont {T.}~\bibnamefont
			{Li}}, \bibinfo {author} {\bibfnamefont {Y.}~\bibnamefont {Li}}, \bibinfo
		{author} {\bibfnamefont {K.~M.}\ \bibnamefont {Liechti}}, \bibinfo {author}
		{\bibfnamefont {N.}~\bibnamefont {Lu}}, \bibinfo {author} {\bibfnamefont
			{H.~S.}\ \bibnamefont {Park}}, \bibinfo {author} {\bibfnamefont {E.~J.}\
			\bibnamefont {Reed}}, \bibinfo {author} {\bibfnamefont {P.}~\bibnamefont
			{Wang}}, \bibinfo {author} {\bibfnamefont {B.~I.}\ \bibnamefont {Yakobson}},
		\bibinfo {author} {\bibfnamefont {T.}~\bibnamefont {Zhang}}, \bibinfo
		{author} {\bibfnamefont {Y.-W.}\ \bibnamefont {Zhang}}, \bibinfo {author}
		{\bibfnamefont {Y.}~\bibnamefont {Zhou}},\ and\ \bibinfo {author}
		{\bibfnamefont {Y.}~\bibnamefont {Zhu}},\ }\bibfield  {title} {\bibinfo
		{title} {A review on mechanics and mechanical properties of 2d
			materials—graphene and beyond},\ }\href
	{https://doi.org/https://doi.org/10.1016/j.eml.2017.01.008} {\bibfield
		{journal} {\bibinfo  {journal} {Extreme Mechanics Letters}\ }\textbf
		{\bibinfo {volume} {13}},\ \bibinfo {pages} {42 } (\bibinfo {year}
		{2017})}\BibitemShut {NoStop}%
	\bibitem [{\citenamefont {Lee}\ \emph {et~al.}(2008)\citenamefont {Lee},
		\citenamefont {Wei}, \citenamefont {Kysar},\ and\ \citenamefont
		{Hone}}]{Lee385}%
	\BibitemOpen
	\bibfield  {author} {\bibinfo {author} {\bibfnamefont {C.}~\bibnamefont
			{Lee}}, \bibinfo {author} {\bibfnamefont {X.}~\bibnamefont {Wei}}, \bibinfo
		{author} {\bibfnamefont {J.~W.}\ \bibnamefont {Kysar}},\ and\ \bibinfo
		{author} {\bibfnamefont {J.}~\bibnamefont {Hone}},\ }\bibfield  {title}
	{\bibinfo {title} {Measurement of the elastic properties and intrinsic
			strength of monolayer graphene},\ }\href
	{https://doi.org/10.1126/science.1157996} {\bibfield  {journal} {\bibinfo
			{journal} {Science}\ }\textbf {\bibinfo {volume} {321}},\ \bibinfo {pages}
		{385} (\bibinfo {year} {2008})}\BibitemShut {NoStop}%
	\bibitem [{\citenamefont {Lindahl}\ \emph {et~al.}(2012)\citenamefont
		{Lindahl}, \citenamefont {Midtvedt}, \citenamefont {Svensson}, \citenamefont
		{Nerushev}, \citenamefont {Lindvall}, \citenamefont {Isacsson},\ and\
		\citenamefont {Campbell}}]{doi:10.1021/nl301080v}%
	\BibitemOpen
	\bibfield  {author} {\bibinfo {author} {\bibfnamefont {N.}~\bibnamefont
			{Lindahl}}, \bibinfo {author} {\bibfnamefont {D.}~\bibnamefont {Midtvedt}},
		\bibinfo {author} {\bibfnamefont {J.}~\bibnamefont {Svensson}}, \bibinfo
		{author} {\bibfnamefont {O.~A.}\ \bibnamefont {Nerushev}}, \bibinfo {author}
		{\bibfnamefont {N.}~\bibnamefont {Lindvall}}, \bibinfo {author}
		{\bibfnamefont {A.}~\bibnamefont {Isacsson}},\ and\ \bibinfo {author}
		{\bibfnamefont {E.~E.~B.}\ \bibnamefont {Campbell}},\ }\bibfield  {title}
	{\bibinfo {title} {Determination of the bending rigidity of graphene via
			electrostatic actuation of buckled membranes},\ }\href@noop {} {\bibfield
		{journal} {\bibinfo  {journal} {Nano Letters}\ }\textbf {\bibinfo {volume}
			{12}},\ \bibinfo {pages} {3526} (\bibinfo {year} {2012})}\BibitemShut
	{NoStop}%
	\bibitem [{\citenamefont {Blees}\ \emph {et~al.}(2015)\citenamefont {Blees},
		\citenamefont {Barnard}, \citenamefont {Rose}, \citenamefont {Roberts},
		\citenamefont {McGill}, \citenamefont {Huang}, \citenamefont {Ruyack},
		\citenamefont {Kevek}, \citenamefont {Kobrin}, \citenamefont {Muller},\ and\
		\citenamefont {McEuen}}]{Blees2015}%
	\BibitemOpen
	\bibfield  {author} {\bibinfo {author} {\bibfnamefont {M.~K.}\ \bibnamefont
			{Blees}}, \bibinfo {author} {\bibfnamefont {A.~W.}\ \bibnamefont {Barnard}},
		\bibinfo {author} {\bibfnamefont {P.~a.}\ \bibnamefont {Rose}}, \bibinfo
		{author} {\bibfnamefont {S.~P.}\ \bibnamefont {Roberts}}, \bibinfo {author}
		{\bibfnamefont {K.~L.}\ \bibnamefont {McGill}}, \bibinfo {author}
		{\bibfnamefont {P.~Y.}\ \bibnamefont {Huang}}, \bibinfo {author}
		{\bibfnamefont {A.~R.}\ \bibnamefont {Ruyack}}, \bibinfo {author}
		{\bibfnamefont {J.~W.}\ \bibnamefont {Kevek}}, \bibinfo {author}
		{\bibfnamefont {B.}~\bibnamefont {Kobrin}}, \bibinfo {author} {\bibfnamefont
			{D.~a.}\ \bibnamefont {Muller}},\ and\ \bibinfo {author} {\bibfnamefont
			{P.~L.}\ \bibnamefont {McEuen}},\ }\bibfield  {title} {\bibinfo {title}
		{{Graphene kirigami}},\ }\href
	{http://www.nature.com/doifinder/10.1038/nature14588} {\bibfield  {journal}
		{\bibinfo  {journal} {Nature}\ }\textbf {\bibinfo {volume} {524}},\ \bibinfo
		{pages} {204} (\bibinfo {year} {2015})}\BibitemShut {NoStop}%
	\bibitem [{\citenamefont {Zalalutdinov}\ \emph {et~al.}(2012)\citenamefont
		{Zalalutdinov}, \citenamefont {Robinson}, \citenamefont {Junkermeier},
		\citenamefont {Culbertson}, \citenamefont {Reinecke}, \citenamefont {Stine},
		\citenamefont {Sheehan}, \citenamefont {Houston},\ and\ \citenamefont
		{Snow}}]{zalalutdinov_NL_12}%
	\BibitemOpen
	\bibfield  {author} {\bibinfo {author} {\bibfnamefont {M.~K.}\ \bibnamefont
			{Zalalutdinov}}, \bibinfo {author} {\bibfnamefont {J.~T.}\ \bibnamefont
			{Robinson}}, \bibinfo {author} {\bibfnamefont {C.~E.}\ \bibnamefont
			{Junkermeier}}, \bibinfo {author} {\bibfnamefont {J.~C.}\ \bibnamefont
			{Culbertson}}, \bibinfo {author} {\bibfnamefont {T.~L.}\ \bibnamefont
			{Reinecke}}, \bibinfo {author} {\bibfnamefont {R.}~\bibnamefont {Stine}},
		\bibinfo {author} {\bibfnamefont {P.~E.}\ \bibnamefont {Sheehan}}, \bibinfo
		{author} {\bibfnamefont {B.~H.}\ \bibnamefont {Houston}},\ and\ \bibinfo
		{author} {\bibfnamefont {E.~S.}\ \bibnamefont {Snow}},\ }\bibfield  {title}
	{\bibinfo {title} {{Engineering graphene mechanical systems.}},\ }\href
	{https://doi.org/10.1021/nl3018059} {\bibfield  {journal} {\bibinfo
			{journal} {Nano letters}\ }\textbf {\bibinfo {volume} {12}},\ \bibinfo
		{pages} {4212} (\bibinfo {year} {2012})}\BibitemShut {NoStop}%
	\bibitem [{\citenamefont {Bunch}\ \emph {et~al.}(2007)\citenamefont {Bunch},
		\citenamefont {van~der Zande}, \citenamefont {Verbridge}, \citenamefont
		{Frank}, \citenamefont {Tanenbaum}, \citenamefont {Parpia}, \citenamefont
		{Craighead},\ and\ \citenamefont {McEuen}}]{Bunch2007}%
	\BibitemOpen
	\bibfield  {author} {\bibinfo {author} {\bibfnamefont {J.~S.}\ \bibnamefont
			{Bunch}}, \bibinfo {author} {\bibfnamefont {A.~M.}\ \bibnamefont {van~der
				Zande}}, \bibinfo {author} {\bibfnamefont {S.~S.}\ \bibnamefont {Verbridge}},
		\bibinfo {author} {\bibfnamefont {I.~W.}\ \bibnamefont {Frank}}, \bibinfo
		{author} {\bibfnamefont {D.~M.}\ \bibnamefont {Tanenbaum}}, \bibinfo {author}
		{\bibfnamefont {J.~M.}\ \bibnamefont {Parpia}}, \bibinfo {author}
		{\bibfnamefont {H.~G.}\ \bibnamefont {Craighead}},\ and\ \bibinfo {author}
		{\bibfnamefont {P.~L.}\ \bibnamefont {McEuen}},\ }\bibfield  {title}
	{\bibinfo {title} {{Electromechanical Resonators from Graphene Sheets}},\
	}\href@noop {} {\bibfield  {journal} {\bibinfo  {journal} {Science}\ }\textbf
		{\bibinfo {volume} {318}},\ \bibinfo {pages} {490} (\bibinfo {year}
		{2007})}\BibitemShut {NoStop}%
	\bibitem [{\citenamefont {Nicholl}\ \emph {et~al.}(2017)\citenamefont
		{Nicholl}, \citenamefont {Lavrik}, \citenamefont {Vlassiouk}, \citenamefont
		{Srijanto},\ and\ \citenamefont {Bolotin}}]{Nicholl2017}%
	\BibitemOpen
	\bibfield  {author} {\bibinfo {author} {\bibfnamefont {R.~J.}\ \bibnamefont
			{Nicholl}}, \bibinfo {author} {\bibfnamefont {N.~V.}\ \bibnamefont {Lavrik}},
		\bibinfo {author} {\bibfnamefont {I.}~\bibnamefont {Vlassiouk}}, \bibinfo
		{author} {\bibfnamefont {B.~R.}\ \bibnamefont {Srijanto}},\ and\ \bibinfo
		{author} {\bibfnamefont {K.~I.}\ \bibnamefont {Bolotin}},\ }\bibfield
	{title} {\bibinfo {title} {{Hidden Area and Mechanical Nonlinearities in
				Freestanding Graphene}},\ }\href
	{https://doi.org/10.1103/PhysRevLett.118.266101} {\bibfield  {journal}
		{\bibinfo  {journal} {Physical Review Letters}\ }\textbf {\bibinfo {volume}
			{118}},\ \bibinfo {pages} {266101} (\bibinfo {year} {2017})}\BibitemShut
	{NoStop}%
	\bibitem [{\citenamefont {Tapaszt{\'{o}}}\ \emph {et~al.}(2012)\citenamefont
		{Tapaszt{\'{o}}}, \citenamefont {Dumitrică}, \citenamefont {Kim},
		\citenamefont {Nemes-Incze}, \citenamefont {Hwang},\ and\ \citenamefont
		{Bir{\'{o}}}}]{tapaszto_nphys_12}%
	\BibitemOpen
	\bibfield  {author} {\bibinfo {author} {\bibfnamefont {L.}~\bibnamefont
			{Tapaszt{\'{o}}}}, \bibinfo {author} {\bibfnamefont {T.}~\bibnamefont
			{Dumitrică}}, \bibinfo {author} {\bibfnamefont {S.~J.}\ \bibnamefont {Kim}},
		\bibinfo {author} {\bibfnamefont {P.}~\bibnamefont {Nemes-Incze}}, \bibinfo
		{author} {\bibfnamefont {C.}~\bibnamefont {Hwang}},\ and\ \bibinfo {author}
		{\bibfnamefont {L.~P.}\ \bibnamefont {Bir{\'{o}}}},\ }\bibfield  {title}
	{\bibinfo {title} {{Breakdown of continuum mechanics for nanometre-wavelength
				rippling of graphene}},\ }\href {https://doi.org/10.1038/nphys2389}
	{\bibfield  {journal} {\bibinfo  {journal} {Nature Physics}\ }\textbf
		{\bibinfo {volume} {8}},\ \bibinfo {pages} {739} (\bibinfo {year}
		{2012})}\BibitemShut {NoStop}%
	\bibitem [{\citenamefont {Chen}\ and\ \citenamefont
		{Chrzan}(2011)}]{chen_PRB_11}%
	\BibitemOpen
	\bibfield  {author} {\bibinfo {author} {\bibfnamefont {S.}~\bibnamefont
			{Chen}}\ and\ \bibinfo {author} {\bibfnamefont {D.~C.}\ \bibnamefont
			{Chrzan}},\ }\bibfield  {title} {\bibinfo {title} {{Monte Carlo simulation of
				temperature-dependent elastic properties of graphene}},\ }\href
	{https://doi.org/10.1103/PhysRevB.84.195409} {\bibfield  {journal} {\bibinfo
			{journal} {Phys. Rev. B}\ }\textbf {\bibinfo {volume} {84}},\ \bibinfo
		{pages} {195409} (\bibinfo {year} {2011})}\BibitemShut {NoStop}%
	\bibitem [{\citenamefont {Neek-Amal}\ and\ \citenamefont
		{Peeters}(2010)}]{neek-amal_PRB_10}%
	\BibitemOpen
	\bibfield  {author} {\bibinfo {author} {\bibfnamefont {M.}~\bibnamefont
			{Neek-Amal}}\ and\ \bibinfo {author} {\bibfnamefont {F.~M.}\ \bibnamefont
			{Peeters}},\ }\bibfield  {title} {\bibinfo {title} {{Graphene nanoribbons
				subjected to axial stress}},\ }\href@noop {} {\bibfield  {journal} {\bibinfo
			{journal} {Phys. Rev. B}\ }\textbf {\bibinfo {volume} {82}},\ \bibinfo
		{pages} {85432} (\bibinfo {year} {2010})}\BibitemShut {NoStop}%
	\bibitem [{\citenamefont {Koskinen}(2014)}]{Koskinen2014}%
	\BibitemOpen
	\bibfield  {author} {\bibinfo {author} {\bibfnamefont {P.}~\bibnamefont
			{Koskinen}},\ }\bibfield  {title} {\bibinfo {title} {{Graphene cardboard:
				From ripples to tunable metamaterial}},\ }\href
	{https://doi.org/10.1063/1.4868125} {\bibfield  {journal} {\bibinfo
			{journal} {Appl. Phys. Lett.}\ }\textbf {\bibinfo {volume} {104}},\ \bibinfo
		{pages} {101902} (\bibinfo {year} {2014})}\BibitemShut {NoStop}%
	\bibitem [{\citenamefont {Wang}\ \emph {et~al.}(2014)\citenamefont {Wang},
		\citenamefont {Liu}, \citenamefont {Li},\ and\ \citenamefont
		{Tan}}]{Wang2014b}%
	\BibitemOpen
	\bibfield  {author} {\bibinfo {author} {\bibfnamefont {C.}~\bibnamefont
			{Wang}}, \bibinfo {author} {\bibfnamefont {Y.}~\bibnamefont {Liu}}, \bibinfo
		{author} {\bibfnamefont {L.}~\bibnamefont {Li}},\ and\ \bibinfo {author}
		{\bibfnamefont {H.}~\bibnamefont {Tan}},\ }\bibfield  {title} {\bibinfo
		{title} {{Anisotropic thermal conductivity of graphene wrinkles}},\ }\href
	{https://doi.org/10.1039/c4nr00423j} {\bibfield  {journal} {\bibinfo
			{journal} {Nanoscale}\ }\textbf {\bibinfo {volume} {6}},\ \bibinfo {pages}
		{5703} (\bibinfo {year} {2014})}\BibitemShut {NoStop}%
	\bibitem [{\citenamefont {Katsnelson}\ and\ \citenamefont
		{Geim}(2008)}]{Katsnelson2008}%
	\BibitemOpen
	\bibfield  {author} {\bibinfo {author} {\bibfnamefont {M.~I.}\ \bibnamefont
			{Katsnelson}}\ and\ \bibinfo {author} {\bibfnamefont {A.~K.}\ \bibnamefont
			{Geim}},\ }\bibfield  {title} {\bibinfo {title} {{Electron scattering on
				microscopic corrugations in graphene}},\ }\href
	{https://doi.org/10.1098/rsta.2007.2157} {\bibfield  {journal} {\bibinfo
			{journal} {Philosophical Transactions of the Royal Society A: Mathematical,
				Physical and Engineering Sciences}\ }\textbf {\bibinfo {volume} {366}},\
		\bibinfo {pages} {195} (\bibinfo {year} {2008})}\BibitemShut {NoStop}%
	\bibitem [{\citenamefont {Kim}\ and\ \citenamefont {{Castro
				Neto}}(2008)}]{kim_EPL_08}%
	\BibitemOpen
	\bibfield  {author} {\bibinfo {author} {\bibfnamefont {E.-A.}\ \bibnamefont
			{Kim}}\ and\ \bibinfo {author} {\bibfnamefont {A.~H.}\ \bibnamefont {{Castro
					Neto}}},\ }\bibfield  {title} {\bibinfo {title} {{Graphene as an electronic
				membrane}},\ }\href@noop {} {\bibfield  {journal} {\bibinfo  {journal} {EPL}\
		}\textbf {\bibinfo {volume} {84}},\ \bibinfo {pages} {57007} (\bibinfo {year}
		{2008})}\BibitemShut {NoStop}%
	\bibitem [{\citenamefont {{Castro Neto}}\ \emph {et~al.}(2009)\citenamefont
		{{Castro Neto}}, \citenamefont {Guinea}, \citenamefont {Peres}, \citenamefont
		{Novoselov},\ and\ \citenamefont {Geim}}]{castro_neto_RMP_09}%
	\BibitemOpen
	\bibfield  {author} {\bibinfo {author} {\bibfnamefont {A.~H.}\ \bibnamefont
			{{Castro Neto}}}, \bibinfo {author} {\bibfnamefont {F.}~\bibnamefont
			{Guinea}}, \bibinfo {author} {\bibfnamefont {N.~M.~R.}\ \bibnamefont
			{Peres}}, \bibinfo {author} {\bibfnamefont {K.~S.}\ \bibnamefont
			{Novoselov}},\ and\ \bibinfo {author} {\bibfnamefont {A.~K.}\ \bibnamefont
			{Geim}},\ }\bibfield  {title} {\bibinfo {title} {{The electronic properties
				of graphene}},\ }\href {https://doi.org/10.1103/RevModPhys.81.109} {\bibfield
		{journal} {\bibinfo  {journal} {Rev. Mod. Phys.}\ }\textbf {\bibinfo
			{volume} {81}},\ \bibinfo {pages} {109} (\bibinfo {year} {2009})}\BibitemShut
	{NoStop}%
	\bibitem [{\citenamefont {{Das Sarma}}\ \emph {et~al.}(2011)\citenamefont {{Das
				Sarma}}, \citenamefont {Adam}, \citenamefont {Hwang},\ and\ \citenamefont
		{Rossi}}]{DasSarma2011}%
	\BibitemOpen
	\bibfield  {author} {\bibinfo {author} {\bibfnamefont {S.}~\bibnamefont {{Das
					Sarma}}}, \bibinfo {author} {\bibfnamefont {S.}~\bibnamefont {Adam}},
		\bibinfo {author} {\bibfnamefont {E.~H.}\ \bibnamefont {Hwang}},\ and\
		\bibinfo {author} {\bibfnamefont {E.}~\bibnamefont {Rossi}},\ }\bibfield
	{title} {\bibinfo {title} {{Electronic transport in two-dimensional
				graphene}},\ }\href {https://doi.org/10.1103/RevModPhys.83.407} {\bibfield
		{journal} {\bibinfo  {journal} {Reviews of Modern Physics}\ }\textbf
		{\bibinfo {volume} {83}},\ \bibinfo {pages} {407} (\bibinfo {year}
		{2011})}\BibitemShut {NoStop}%
	\bibitem [{\citenamefont {Wang}\ \emph {et~al.}(2016)\citenamefont {Wang},
		\citenamefont {Gao},\ and\ \citenamefont {Huang}}]{Wang2015a}%
	\BibitemOpen
	\bibfield  {author} {\bibinfo {author} {\bibfnamefont {P.}~\bibnamefont
			{Wang}}, \bibinfo {author} {\bibfnamefont {W.}~\bibnamefont {Gao}},\ and\
		\bibinfo {author} {\bibfnamefont {R.}~\bibnamefont {Huang}},\ }\bibfield
	{title} {\bibinfo {title} {{Entropic Effects of Thermal Rippling on van der
				Waals Interactions between Monolayer Graphene and a Rigid Substrate}},\
	}\href {http://arxiv.org/abs/1511.02914} {\bibfield  {journal} {\bibinfo
			{journal} {J. Appl. Phys.}\ }\textbf {\bibinfo {volume} {119}},\ \bibinfo
		{pages} {074305} (\bibinfo {year} {2016})},\ \Eprint
	{https://arxiv.org/abs/1511.02914} {arXiv:1511.02914} \BibitemShut {NoStop}%
	\bibitem [{\citenamefont {He}\ \emph {et~al.}(2013)\citenamefont {He},
		\citenamefont {Chen}, \citenamefont {Yu}, \citenamefont {Ouyang},\ and\
		\citenamefont {Yang}}]{He2013a}%
	\BibitemOpen
	\bibfield  {author} {\bibinfo {author} {\bibfnamefont {Y.}~\bibnamefont
			{He}}, \bibinfo {author} {\bibfnamefont {W.~F.}\ \bibnamefont {Chen}},
		\bibinfo {author} {\bibfnamefont {W.~B.}\ \bibnamefont {Yu}}, \bibinfo
		{author} {\bibfnamefont {G.}~\bibnamefont {Ouyang}},\ and\ \bibinfo {author}
		{\bibfnamefont {G.~W.}\ \bibnamefont {Yang}},\ }\bibfield  {title} {\bibinfo
		{title} {{Anomalous interface adhesion of graphene membranes.}},\ }\href
	{https://doi.org/10.1038/srep02660} {\bibfield  {journal} {\bibinfo
			{journal} {Scientific reports}\ }\textbf {\bibinfo {volume} {3}},\ \bibinfo
		{pages} {2660} (\bibinfo {year} {2013})}\BibitemShut {NoStop}%
	\bibitem [{\citenamefont {Gao}\ \emph {et~al.}(2015)\citenamefont {Gao},
		\citenamefont {Kim}, \citenamefont {Zhou}, \citenamefont {Chiu},
		\citenamefont {N{\'{e}}lias}, \citenamefont {Berger}, \citenamefont
		{de~Heer}, \citenamefont {Polloni}, \citenamefont {Sordan}, \citenamefont
		{Bongiorno},\ and\ \citenamefont {Riedo}}]{Gao2015a}%
	\BibitemOpen
	\bibfield  {author} {\bibinfo {author} {\bibfnamefont {Y.}~\bibnamefont
			{Gao}}, \bibinfo {author} {\bibfnamefont {S.}~\bibnamefont {Kim}}, \bibinfo
		{author} {\bibfnamefont {S.}~\bibnamefont {Zhou}}, \bibinfo {author}
		{\bibfnamefont {H.-C.}\ \bibnamefont {Chiu}}, \bibinfo {author}
		{\bibfnamefont {D.}~\bibnamefont {N{\'{e}}lias}}, \bibinfo {author}
		{\bibfnamefont {C.}~\bibnamefont {Berger}}, \bibinfo {author} {\bibfnamefont
			{W.}~\bibnamefont {de~Heer}}, \bibinfo {author} {\bibfnamefont
			{L.}~\bibnamefont {Polloni}}, \bibinfo {author} {\bibfnamefont
			{R.}~\bibnamefont {Sordan}}, \bibinfo {author} {\bibfnamefont
			{A.}~\bibnamefont {Bongiorno}},\ and\ \bibinfo {author} {\bibfnamefont
			{E.}~\bibnamefont {Riedo}},\ }\bibfield  {title} {\bibinfo {title} {{Elastic
				coupling between layers in two-dimensional materials.}},\ }\href
	{https://doi.org/10.1038/nmat4322} {\bibfield  {journal} {\bibinfo  {journal}
			{Nature materials}\ }\textbf {\bibinfo {volume} {14}},\ \bibinfo {pages}
		{714} (\bibinfo {year} {2015})}\BibitemShut {NoStop}%
	\bibitem [{\citenamefont {Wang}\ \emph {et~al.}(2012)\citenamefont {Wang},
		\citenamefont {Bhandari}, \citenamefont {Yi}, \citenamefont {Bell},
		\citenamefont {Westervelt},\ and\ \citenamefont {Kaxiras}}]{Wang2012b}%
	\BibitemOpen
	\bibfield  {author} {\bibinfo {author} {\bibfnamefont {W.~L.}\ \bibnamefont
			{Wang}}, \bibinfo {author} {\bibfnamefont {S.}~\bibnamefont {Bhandari}},
		\bibinfo {author} {\bibfnamefont {W.}~\bibnamefont {Yi}}, \bibinfo {author}
		{\bibfnamefont {D.~C.}\ \bibnamefont {Bell}}, \bibinfo {author}
		{\bibfnamefont {R.}~\bibnamefont {Westervelt}},\ and\ \bibinfo {author}
		{\bibfnamefont {E.}~\bibnamefont {Kaxiras}},\ }\bibfield  {title} {\bibinfo
		{title} {{Direct imaging of atomic-scale ripples in few-layer graphene}},\
	}\href {https://doi.org/10.1021/nl300071y} {\bibfield  {journal} {\bibinfo
			{journal} {Nano Letters}\ }\textbf {\bibinfo {volume} {12}},\ \bibinfo
		{pages} {2278} (\bibinfo {year} {2012})}\BibitemShut {NoStop}%
	\bibitem [{\citenamefont {Deng}\ and\ \citenamefont {Berry}(2016)}]{Deng2016a}%
	\BibitemOpen
	\bibfield  {author} {\bibinfo {author} {\bibfnamefont {S.}~\bibnamefont
			{Deng}}\ and\ \bibinfo {author} {\bibfnamefont {V.}~\bibnamefont {Berry}},\
	}\bibfield  {title} {\bibinfo {title} {{Wrinkled, rippled and crumpled
				graphene: An overview of formation mechanism, electronic properties, and
				applications}},\ }\href {https://doi.org/10.1016/j.mattod.2015.10.002}
	{\bibfield  {journal} {\bibinfo  {journal} {Materials Today}\ }\textbf
		{\bibinfo {volume} {19}},\ \bibinfo {pages} {197} (\bibinfo {year}
		{2016})}\BibitemShut {NoStop}%
	\bibitem [{\citenamefont {Thompson-Flagg}\ \emph {et~al.}(2009)\citenamefont
		{Thompson-Flagg}, \citenamefont {Moura},\ and\ \citenamefont
		{Marder}}]{thompson-flagg_EPL_09}%
	\BibitemOpen
	\bibfield  {author} {\bibinfo {author} {\bibfnamefont {R.~C.}\ \bibnamefont
			{Thompson-Flagg}}, \bibinfo {author} {\bibfnamefont {M.~J.~B.}\ \bibnamefont
			{Moura}},\ and\ \bibinfo {author} {\bibfnamefont {M.}~\bibnamefont
			{Marder}},\ }\bibfield  {title} {\bibinfo {title} {{Rippling of graphene}},\
	}\href@noop {} {\bibfield  {journal} {\bibinfo  {journal} {EPL}\ }\textbf
		{\bibinfo {volume} {85}},\ \bibinfo {pages} {46002} (\bibinfo {year}
		{2009})}\BibitemShut {NoStop}%
	\bibitem [{\citenamefont {Gao}\ and\ \citenamefont {Huang}(2014)}]{GAO201442}%
	\BibitemOpen
	\bibfield  {author} {\bibinfo {author} {\bibfnamefont {W.}~\bibnamefont
			{Gao}}\ and\ \bibinfo {author} {\bibfnamefont {R.}~\bibnamefont {Huang}},\
	}\bibfield  {title} {\bibinfo {title} {Thermomechanics of monolayer graphene:
			Rippling, thermal expansion and elasticity},\ }\href
	{https://doi.org/https://doi.org/10.1016/j.jmps.2014.01.011} {\bibfield
		{journal} {\bibinfo  {journal} {Journal of the Mechanics and Physics of
				Solids}\ }\textbf {\bibinfo {volume} {66}},\ \bibinfo {pages} {42 } (\bibinfo
		{year} {2014})}\BibitemShut {NoStop}%
	\bibitem [{\citenamefont {Kholmanov}\ \emph {et~al.}(2009)\citenamefont
		{Kholmanov}, \citenamefont {Cavaliere}, \citenamefont {Fanetti},
		\citenamefont {Cepek},\ and\ \citenamefont {Gavioli}}]{kholmanov_PRB_09}%
	\BibitemOpen
	\bibfield  {author} {\bibinfo {author} {\bibfnamefont {I.}~\bibnamefont
			{Kholmanov}}, \bibinfo {author} {\bibfnamefont {E.}~\bibnamefont
			{Cavaliere}}, \bibinfo {author} {\bibfnamefont {M.}~\bibnamefont {Fanetti}},
		\bibinfo {author} {\bibfnamefont {C.}~\bibnamefont {Cepek}},\ and\ \bibinfo
		{author} {\bibfnamefont {L.}~\bibnamefont {Gavioli}},\ }\bibfield  {title}
	{\bibinfo {title} {{Growth of curved graphene sheets on graphite by chemical
				vapor deposition}},\ }\href {https://doi.org/10.1103/PhysRevB.79.233403}
	{\bibfield  {journal} {\bibinfo  {journal} {Physical Review B}\ }\textbf
		{\bibinfo {volume} {79}},\ \bibinfo {pages} {233403} (\bibinfo {year}
		{2009})}\BibitemShut {NoStop}%
	\bibitem [{\citenamefont {Ko\ifmmode~\check{s}\else \v{s}\fi{}mrlj}\ and\
		\citenamefont {Nelson}(2016)}]{PhysRevB.93.125431}%
	\BibitemOpen
	\bibfield  {author} {\bibinfo {author} {\bibfnamefont {A.}~\bibnamefont
			{Ko\ifmmode~\check{s}\else \v{s}\fi{}mrlj}}\ and\ \bibinfo {author}
		{\bibfnamefont {D.~R.}\ \bibnamefont {Nelson}},\ }\bibfield  {title}
	{\bibinfo {title} {Response of thermalized ribbons to pulling and bending},\
	}\href {https://doi.org/10.1103/PhysRevB.93.125431} {\bibfield  {journal}
		{\bibinfo  {journal} {Phys. Rev. B}\ }\textbf {\bibinfo {volume} {93}},\
		\bibinfo {pages} {125431} (\bibinfo {year} {2016})}\BibitemShut {NoStop}%
	\bibitem [{\citenamefont {Banhart}\ \emph {et~al.}(2011)\citenamefont
		{Banhart}, \citenamefont {Kotakoski},\ and\ \citenamefont
		{Krasheninnikov}}]{doi:10.1021/nn102598m}%
	\BibitemOpen
	\bibfield  {author} {\bibinfo {author} {\bibfnamefont {F.}~\bibnamefont
			{Banhart}}, \bibinfo {author} {\bibfnamefont {J.}~\bibnamefont {Kotakoski}},\
		and\ \bibinfo {author} {\bibfnamefont {A.~V.}\ \bibnamefont
			{Krasheninnikov}},\ }\bibfield  {title} {\bibinfo {title} {Structural defects
			in graphene},\ }\href@noop {} {\bibfield  {journal} {\bibinfo  {journal} {ACS
				Nano}\ }\textbf {\bibinfo {volume} {5}},\ \bibinfo {pages} {26} (\bibinfo
		{year} {2011})}\BibitemShut {NoStop}%
	\bibitem [{\citenamefont {Lee}\ \emph {et~al.}(2013)\citenamefont {Lee},
		\citenamefont {Cooper}, \citenamefont {An}, \citenamefont {Lee},
		\citenamefont {van~der Zande}, \citenamefont {Petrone}, \citenamefont
		{Hammerberg}, \citenamefont {Lee}, \citenamefont {Crawford}, \citenamefont
		{Oliver}, \citenamefont {Kysar},\ and\ \citenamefont {Hone}}]{Lee1073}%
	\BibitemOpen
	\bibfield  {author} {\bibinfo {author} {\bibfnamefont {G.-H.}\ \bibnamefont
			{Lee}}, \bibinfo {author} {\bibfnamefont {R.~C.}\ \bibnamefont {Cooper}},
		\bibinfo {author} {\bibfnamefont {S.~J.}\ \bibnamefont {An}}, \bibinfo
		{author} {\bibfnamefont {S.}~\bibnamefont {Lee}}, \bibinfo {author}
		{\bibfnamefont {A.}~\bibnamefont {van~der Zande}}, \bibinfo {author}
		{\bibfnamefont {N.}~\bibnamefont {Petrone}}, \bibinfo {author} {\bibfnamefont
			{A.~G.}\ \bibnamefont {Hammerberg}}, \bibinfo {author} {\bibfnamefont
			{C.}~\bibnamefont {Lee}}, \bibinfo {author} {\bibfnamefont {B.}~\bibnamefont
			{Crawford}}, \bibinfo {author} {\bibfnamefont {W.}~\bibnamefont {Oliver}},
		\bibinfo {author} {\bibfnamefont {J.~W.}\ \bibnamefont {Kysar}},\ and\
		\bibinfo {author} {\bibfnamefont {J.}~\bibnamefont {Hone}},\ }\bibfield
	{title} {\bibinfo {title} {High-strength chemical-vapor{\textendash}deposited
			graphene and grain boundaries},\ }\href
	{https://doi.org/10.1126/science.1235126} {\bibfield  {journal} {\bibinfo
			{journal} {Science}\ }\textbf {\bibinfo {volume} {340}},\ \bibinfo {pages}
		{1073} (\bibinfo {year} {2013})}\BibitemShut {NoStop}%
	\bibitem [{\citenamefont {Bao}\ \emph {et~al.}(2009)\citenamefont {Bao},
		\citenamefont {Miao}, \citenamefont {Chen}, \citenamefont {Zhang},
		\citenamefont {Jang}, \citenamefont {Dames},\ and\ \citenamefont
		{Lau}}]{Bao2009}%
	\BibitemOpen
	\bibfield  {author} {\bibinfo {author} {\bibfnamefont {W.}~\bibnamefont
			{Bao}}, \bibinfo {author} {\bibfnamefont {F.}~\bibnamefont {Miao}}, \bibinfo
		{author} {\bibfnamefont {Z.}~\bibnamefont {Chen}}, \bibinfo {author}
		{\bibfnamefont {H.}~\bibnamefont {Zhang}}, \bibinfo {author} {\bibfnamefont
			{W.}~\bibnamefont {Jang}}, \bibinfo {author} {\bibfnamefont {C.}~\bibnamefont
			{Dames}},\ and\ \bibinfo {author} {\bibfnamefont {C.~N.}\ \bibnamefont
			{Lau}},\ }\bibfield  {title} {\bibinfo {title} {{Controlled ripple texturing
				of suspended graphene and ultrathin graphite membranes.}},\ }\href
	{https://doi.org/10.1038/nnano.2009.191} {\bibfield  {journal} {\bibinfo
			{journal} {Nat. Nanotechnol.}\ }\textbf {\bibinfo {volume} {4}},\ \bibinfo
		{pages} {562} (\bibinfo {year} {2009})}\BibitemShut {NoStop}%
	\bibitem [{\citenamefont {Liu}\ \emph {et~al.}(2011)\citenamefont {Liu},
		\citenamefont {Pan}, \citenamefont {Fu}, \citenamefont {Zhang},\ and\
		\citenamefont {Dai}}]{Liu2011}%
	\BibitemOpen
	\bibfield  {author} {\bibinfo {author} {\bibfnamefont {N.}~\bibnamefont
			{Liu}}, \bibinfo {author} {\bibfnamefont {Z.}~\bibnamefont {Pan}}, \bibinfo
		{author} {\bibfnamefont {L.}~\bibnamefont {Fu}}, \bibinfo {author}
		{\bibfnamefont {C.}~\bibnamefont {Zhang}},\ and\ \bibinfo {author}
		{\bibfnamefont {B.}~\bibnamefont {Dai}},\ }\bibfield  {title} {\bibinfo
		{title} {{The origin of wrinkles on transferred graphene}},\ }\href
	{https://doi.org/10.1007/s12274-011-0156-3} {\bibfield  {journal} {\bibinfo
			{journal} {Nano Research}\ }\textbf {\bibinfo {volume} {4}},\ \bibinfo
		{pages} {996} (\bibinfo {year} {2011})}\BibitemShut {NoStop}%
	\bibitem [{\citenamefont {Duan}\ \emph {et~al.}(2011)\citenamefont {Duan},
		\citenamefont {Gong},\ and\ \citenamefont {Wang}}]{Duan2011}%
	\BibitemOpen
	\bibfield  {author} {\bibinfo {author} {\bibfnamefont {W.~H.}\ \bibnamefont
			{Duan}}, \bibinfo {author} {\bibfnamefont {K.}~\bibnamefont {Gong}},\ and\
		\bibinfo {author} {\bibfnamefont {Q.}~\bibnamefont {Wang}},\ }\bibfield
	{title} {\bibinfo {title} {{Controlling the formation of wrinkles in a single
				layer graphene sheet subjected to in-plane shear}},\ }\href
	{https://doi.org/10.1016/j.carbon.2011.03.033} {\bibfield  {journal}
		{\bibinfo  {journal} {Carbon}\ }\textbf {\bibinfo {volume} {49}},\ \bibinfo
		{pages} {3107} (\bibinfo {year} {2011})}\BibitemShut {NoStop}%
	\bibitem [{\citenamefont {Ludacka}\ \emph {et~al.}(2018)\citenamefont
		{Ludacka}, \citenamefont {Monazam}, \citenamefont {Rentenberger},
		\citenamefont {Friedrich}, \citenamefont {Stefanelli}, \citenamefont
		{Meyer},\ and\ \citenamefont {Kotakoski}}]{Ludacka2018}%
	\BibitemOpen
	\bibfield  {author} {\bibinfo {author} {\bibfnamefont {U.}~\bibnamefont
			{Ludacka}}, \bibinfo {author} {\bibfnamefont {M.~R.}\ \bibnamefont
			{Monazam}}, \bibinfo {author} {\bibfnamefont {C.}~\bibnamefont
			{Rentenberger}}, \bibinfo {author} {\bibfnamefont {M.}~\bibnamefont
			{Friedrich}}, \bibinfo {author} {\bibfnamefont {U.}~\bibnamefont
			{Stefanelli}}, \bibinfo {author} {\bibfnamefont {J.~C.}\ \bibnamefont
			{Meyer}},\ and\ \bibinfo {author} {\bibfnamefont {J.}~\bibnamefont
			{Kotakoski}},\ }\bibfield  {title} {\bibinfo {title} {{In situ control of
				graphene ripples and strain in the electron microscope}},\ }\href
	{https://doi.org/10.1038/s41699-018-0069-z} {\bibfield  {journal} {\bibinfo
			{journal} {npj 2D Materials and Applications}\ }\textbf {\bibinfo {volume}
			{2}},\ \bibinfo {pages} {1} (\bibinfo {year} {2018})}\BibitemShut {NoStop}%
	\bibitem [{\citenamefont {Johansson}\ \emph {et~al.}(2017)\citenamefont
		{Johansson}, \citenamefont {Myllyperki{\"{o}}}, \citenamefont {Koskinen},
		\citenamefont {Aumanen}, \citenamefont {Koivistoinen}, \citenamefont {Tsai},
		\citenamefont {Chen}, \citenamefont {Chang}, \citenamefont {Hiltunen},
		\citenamefont {Manninen}, \citenamefont {Woon},\ and\ \citenamefont
		{Pettersson}}]{Johansson2017}%
	\BibitemOpen
	\bibfield  {author} {\bibinfo {author} {\bibfnamefont {A.}~\bibnamefont
			{Johansson}}, \bibinfo {author} {\bibfnamefont {P.}~\bibnamefont
			{Myllyperki{\"{o}}}}, \bibinfo {author} {\bibfnamefont {P.}~\bibnamefont
			{Koskinen}}, \bibinfo {author} {\bibfnamefont {J.}~\bibnamefont {Aumanen}},
		\bibinfo {author} {\bibfnamefont {J.}~\bibnamefont {Koivistoinen}}, \bibinfo
		{author} {\bibfnamefont {H.~C.}\ \bibnamefont {Tsai}}, \bibinfo {author}
		{\bibfnamefont {C.~H.}\ \bibnamefont {Chen}}, \bibinfo {author}
		{\bibfnamefont {L.~Y.}\ \bibnamefont {Chang}}, \bibinfo {author}
		{\bibfnamefont {V.~M.}\ \bibnamefont {Hiltunen}}, \bibinfo {author}
		{\bibfnamefont {J.~J.}\ \bibnamefont {Manninen}}, \bibinfo {author}
		{\bibfnamefont {W.~Y.}\ \bibnamefont {Woon}},\ and\ \bibinfo {author}
		{\bibfnamefont {M.}~\bibnamefont {Pettersson}},\ }\bibfield  {title}
	{\bibinfo {title} {{Optical Forging of Graphene into Three-Dimensional
				Shapes}},\ }\href {https://doi.org/10.1021/acs.nanolett.7b03530} {\bibfield
		{journal} {\bibinfo  {journal} {Nano Lett.}\ }\textbf {\bibinfo {volume}
			{17}},\ \bibinfo {pages} {6469} (\bibinfo {year} {2017})}\BibitemShut
	{NoStop}%
	\bibitem [{\citenamefont {Koskinen}\ \emph {et~al.}(2018)\citenamefont
		{Koskinen}, \citenamefont {Karppinen}, \citenamefont {Myllyperki{\"{o}}},
		\citenamefont {Hiltunen}, \citenamefont {Johansson},\ and\ \citenamefont
		{Pettersson}}]{Koskinen2018a}%
	\BibitemOpen
	\bibfield  {author} {\bibinfo {author} {\bibfnamefont {P.}~\bibnamefont
			{Koskinen}}, \bibinfo {author} {\bibfnamefont {K.}~\bibnamefont {Karppinen}},
		\bibinfo {author} {\bibfnamefont {P.}~\bibnamefont {Myllyperki{\"{o}}}},
		\bibinfo {author} {\bibfnamefont {V.-M.}\ \bibnamefont {Hiltunen}}, \bibinfo
		{author} {\bibfnamefont {A.}~\bibnamefont {Johansson}},\ and\ \bibinfo
		{author} {\bibfnamefont {M.}~\bibnamefont {Pettersson}},\ }\bibfield  {title}
	{\bibinfo {title} {{Optically Forged Diffraction-Unlimited Ripples in
				Graphene}},\ }\href {https://doi.org/10.1021/acs.jpclett.8b02461} {\bibfield
		{journal} {\bibinfo  {journal} {The Journal of Physical Chemistry Letters}\
			,\ \bibinfo {pages} {6179}} (\bibinfo {year} {2018})}\BibitemShut {NoStop}%
	\bibitem [{\citenamefont {Koivistoinen}\ \emph {et~al.}(2016)\citenamefont
		{Koivistoinen}, \citenamefont {Sladkova}, \citenamefont {Aumanen},
		\citenamefont {Koskinen}, \citenamefont {Roberts}, \citenamefont {Johansson},
		\citenamefont {Myllyperki{\"{o}}},\ and\ \citenamefont
		{Pettersson}}]{Koivistoinen2016}%
	\BibitemOpen
	\bibfield  {author} {\bibinfo {author} {\bibfnamefont {J.}~\bibnamefont
			{Koivistoinen}}, \bibinfo {author} {\bibfnamefont {L.}~\bibnamefont
			{Sladkova}}, \bibinfo {author} {\bibfnamefont {J.}~\bibnamefont {Aumanen}},
		\bibinfo {author} {\bibfnamefont {P.~J.}\ \bibnamefont {Koskinen}}, \bibinfo
		{author} {\bibfnamefont {K.}~\bibnamefont {Roberts}}, \bibinfo {author}
		{\bibfnamefont {A.}~\bibnamefont {Johansson}}, \bibinfo {author}
		{\bibfnamefont {P.}~\bibnamefont {Myllyperki{\"{o}}}},\ and\ \bibinfo
		{author} {\bibfnamefont {M.}~\bibnamefont {Pettersson}},\ }\bibfield  {title}
	{\bibinfo {title} {{From Seeds to Islands: Growth of Oxidized Graphene by
				Two-Photon Oxidation}},\ }\href {https://doi.org/10.1021/acs.jpcc.6b06099}
	{\bibfield  {journal} {\bibinfo  {journal} {J. Phys. Chem. C}\ }\textbf
		{\bibinfo {volume} {120}},\ \bibinfo {pages} {22330} (\bibinfo {year}
		{2016})}\BibitemShut {NoStop}%
	\bibitem [{\citenamefont {Hiltunen}\ \emph {et~al.}(2020)\citenamefont
		{Hiltunen}, \citenamefont {Koskinen}, \citenamefont {Mentel}, \citenamefont
		{Manninen}, \citenamefont {Myllyperki{\"{o}}}, \citenamefont {Johansson},\
		and\ \citenamefont {Pettersson}}]{Hiltunen2020}%
	\BibitemOpen
	\bibfield  {author} {\bibinfo {author} {\bibfnamefont {V.-M.}\ \bibnamefont
			{Hiltunen}}, \bibinfo {author} {\bibfnamefont {P.~J.}\ \bibnamefont
			{Koskinen}}, \bibinfo {author} {\bibfnamefont {K.~K.}\ \bibnamefont
			{Mentel}}, \bibinfo {author} {\bibfnamefont {J.}~\bibnamefont {Manninen}},
		\bibinfo {author} {\bibfnamefont {P.}~\bibnamefont {Myllyperki{\"{o}}}},
		\bibinfo {author} {\bibfnamefont {A.}~\bibnamefont {Johansson}},\ and\
		\bibinfo {author} {\bibfnamefont {M.}~\bibnamefont {Pettersson}},\ }\bibfield
	{title} {\bibinfo {title} {{Making Graphene Luminescent by Direct Laser
				Writing}},\ }\href {https://doi.org/10.1021/acs.jpcc.0c00194} {\bibfield
		{journal} {\bibinfo  {journal} {The Journal of Physical Chemistry C}\
		}\textbf {\bibinfo {volume} {124}},\ \bibinfo {pages} {8371} (\bibinfo {year}
		{2020})}\BibitemShut {NoStop}%
	\bibitem [{\citenamefont {Ko\ifmmode~\check{s}\else \v{s}\fi{}mrlj}\ and\
		\citenamefont {Nelson}(2013)}]{PhysRevE.88.012136}%
	\BibitemOpen
	\bibfield  {author} {\bibinfo {author} {\bibfnamefont {A.}~\bibnamefont
			{Ko\ifmmode~\check{s}\else \v{s}\fi{}mrlj}}\ and\ \bibinfo {author}
		{\bibfnamefont {D.~R.}\ \bibnamefont {Nelson}},\ }\bibfield  {title}
	{\bibinfo {title} {Mechanical properties of warped membranes},\ }\href
	{https://doi.org/10.1103/PhysRevE.88.012136} {\bibfield  {journal} {\bibinfo
			{journal} {Phys. Rev. E}\ }\textbf {\bibinfo {volume} {88}},\ \bibinfo
		{pages} {012136} (\bibinfo {year} {2013})}\BibitemShut {NoStop}%
	\bibitem [{\citenamefont {Liu}\ and\ \citenamefont
		{Yakobson}(2010)}]{Liu2010a}%
	\BibitemOpen
	\bibfield  {author} {\bibinfo {author} {\bibfnamefont {Y.}~\bibnamefont
			{Liu}}\ and\ \bibinfo {author} {\bibfnamefont {B.~I.}\ \bibnamefont
			{Yakobson}},\ }\bibfield  {title} {\bibinfo {title} {{Cones, pringles, and
				grain boundary landscapes in graphene topology.}},\ }\href
	{https://doi.org/10.1021/nl100988r} {\bibfield  {journal} {\bibinfo
			{journal} {Nano letters}\ }\textbf {\bibinfo {volume} {10}},\ \bibinfo
		{pages} {2178} (\bibinfo {year} {2010})}\BibitemShut {NoStop}%
	\bibitem [{\citenamefont {Malola}\ \emph {et~al.}(2010)\citenamefont {Malola},
		\citenamefont {H{\"{a}}kkinen},\ and\ \citenamefont
		{Koskinen}}]{malola_PRB_10}%
	\BibitemOpen
	\bibfield  {author} {\bibinfo {author} {\bibfnamefont {S.}~\bibnamefont
			{Malola}}, \bibinfo {author} {\bibfnamefont {H.}~\bibnamefont
			{H{\"{a}}kkinen}},\ and\ \bibinfo {author} {\bibfnamefont {P.}~\bibnamefont
			{Koskinen}},\ }\bibfield  {title} {\bibinfo {title} {{Structural, chemical,
				and dynamical trends in graphene grain boundaries}},\ }\href@noop {}
	{\bibfield  {journal} {\bibinfo  {journal} {Phys. Rev. B}\ }\textbf {\bibinfo
			{volume} {81}},\ \bibinfo {pages} {165447} (\bibinfo {year}
		{2010})}\BibitemShut {NoStop}%
	\bibitem [{\citenamefont {Ma}\ \emph {et~al.}(2009)\citenamefont {Ma},
		\citenamefont {Alf{\`{e}}}, \citenamefont {Michaelides},\ and\ \citenamefont
		{Wang}}]{Ma2009}%
	\BibitemOpen
	\bibfield  {author} {\bibinfo {author} {\bibfnamefont {J.}~\bibnamefont
			{Ma}}, \bibinfo {author} {\bibfnamefont {D.}~\bibnamefont {Alf{\`{e}}}},
		\bibinfo {author} {\bibfnamefont {A.}~\bibnamefont {Michaelides}},\ and\
		\bibinfo {author} {\bibfnamefont {E.}~\bibnamefont {Wang}},\ }\bibfield
	{title} {\bibinfo {title} {{Stone-Wales defects in graphene and other planar
				s p2 -bonded materials}},\ }\href
	{https://doi.org/10.1103/PhysRevB.80.033407} {\bibfield  {journal} {\bibinfo
			{journal} {Physical Review B - Condensed Matter and Materials Physics}\
		}\textbf {\bibinfo {volume} {80}},\ \bibinfo {pages} {033407} (\bibinfo
		{year} {2009})}\BibitemShut {NoStop}%
	\bibitem [{\citenamefont {Fan}\ \emph {et~al.}(2010)\citenamefont {Fan},
		\citenamefont {Yang},\ and\ \citenamefont {Zhang}}]{Fan2010}%
	\BibitemOpen
	\bibfield  {author} {\bibinfo {author} {\bibfnamefont {B.~B.}\ \bibnamefont
			{Fan}}, \bibinfo {author} {\bibfnamefont {X.~B.}\ \bibnamefont {Yang}},\ and\
		\bibinfo {author} {\bibfnamefont {R.}~\bibnamefont {Zhang}},\ }\bibfield
	{title} {\bibinfo {title} {{Anisotropic mechanical properties and Stone-Wales
				defects in graphene monolayer: A theoretical study}},\ }\href
	{https://doi.org/10.1016/j.physleta.2010.04.066} {\bibfield  {journal}
		{\bibinfo  {journal} {Physics Letters, Section A: General, Atomic and Solid
				State Physics}\ }\textbf {\bibinfo {volume} {374}},\ \bibinfo {pages} {2781}
		(\bibinfo {year} {2010})}\BibitemShut {NoStop}%
	\bibitem [{\citenamefont {Porezag}\ \emph {et~al.}(1995)\citenamefont
		{Porezag}, \citenamefont {Frauenheim}, \citenamefont {K{\"{o}}hler},
		\citenamefont {Seifert},\ and\ \citenamefont {Kaschner}}]{porezag_PRB_95}%
	\BibitemOpen
	\bibfield  {author} {\bibinfo {author} {\bibfnamefont {D.}~\bibnamefont
			{Porezag}}, \bibinfo {author} {\bibfnamefont {T.}~\bibnamefont {Frauenheim}},
		\bibinfo {author} {\bibfnamefont {T.}~\bibnamefont {K{\"{o}}hler}}, \bibinfo
		{author} {\bibfnamefont {G.}~\bibnamefont {Seifert}},\ and\ \bibinfo {author}
		{\bibfnamefont {R.}~\bibnamefont {Kaschner}},\ }\bibfield  {title} {\bibinfo
		{title} {{Construction of tight-binding-like potentials on the basis of
				density-functional theory: application to carbon}},\ }\href@noop {}
	{\bibfield  {journal} {\bibinfo  {journal} {Phys. Rev. B}\ }\textbf {\bibinfo
			{volume} {51}},\ \bibinfo {pages} {12947} (\bibinfo {year}
		{1995})}\BibitemShut {NoStop}%
	\bibitem [{\citenamefont {Elstner}\ \emph {et~al.}(1998)\citenamefont
		{Elstner}, \citenamefont {Porezag}, \citenamefont {Jungnickel}, \citenamefont
		{Elsner}, \citenamefont {Haugk}, \citenamefont {Frauenheim}, \citenamefont
		{Suhai},\ and\ \citenamefont {Seifert}}]{elstner_PRB_98}%
	\BibitemOpen
	\bibfield  {author} {\bibinfo {author} {\bibfnamefont {M.}~\bibnamefont
			{Elstner}}, \bibinfo {author} {\bibfnamefont {D.}~\bibnamefont {Porezag}},
		\bibinfo {author} {\bibfnamefont {G.}~\bibnamefont {Jungnickel}}, \bibinfo
		{author} {\bibfnamefont {J.}~\bibnamefont {Elsner}}, \bibinfo {author}
		{\bibfnamefont {M.}~\bibnamefont {Haugk}}, \bibinfo {author} {\bibfnamefont
			{T.}~\bibnamefont {Frauenheim}}, \bibinfo {author} {\bibfnamefont
			{S.}~\bibnamefont {Suhai}},\ and\ \bibinfo {author} {\bibfnamefont
			{G.}~\bibnamefont {Seifert}},\ }\bibfield  {title} {\bibinfo {title}
		{{Self-consistent-charge density-functional tight-binding method for
				simulations of complex materials properties}},\ }\href@noop {} {\bibfield
		{journal} {\bibinfo  {journal} {Phys. Rev. B}\ }\textbf {\bibinfo {volume}
			{58}},\ \bibinfo {pages} {7260} (\bibinfo {year} {1998})}\BibitemShut
	{NoStop}%
	\bibitem [{\citenamefont {Frauenheim}\ \emph {et~al.}(2000)\citenamefont
		{Frauenheim}, \citenamefont {Seifert}, \citenamefont {Elstner}, \citenamefont
		{Hajnal}, \citenamefont {Jungnickel}, \citenamefont {Porezag}, \citenamefont
		{Suhai},\ and\ \citenamefont {Scholz}}]{frauenheim_PSSb_00}%
	\BibitemOpen
	\bibfield  {author} {\bibinfo {author} {\bibfnamefont {T.}~\bibnamefont
			{Frauenheim}}, \bibinfo {author} {\bibfnamefont {G.}~\bibnamefont {Seifert}},
		\bibinfo {author} {\bibfnamefont {M.}~\bibnamefont {Elstner}}, \bibinfo
		{author} {\bibfnamefont {Z.}~\bibnamefont {Hajnal}}, \bibinfo {author}
		{\bibfnamefont {G.}~\bibnamefont {Jungnickel}}, \bibinfo {author}
		{\bibfnamefont {D.}~\bibnamefont {Porezag}}, \bibinfo {author} {\bibfnamefont
			{S.}~\bibnamefont {Suhai}},\ and\ \bibinfo {author} {\bibfnamefont
			{R.}~\bibnamefont {Scholz}},\ }\bibfield  {title} {\bibinfo {title} {{A
				Self-Consistent Charge Density-Functional Based Tight-Binding Method for
				Predictive Materials Simulations in Physics, Chemistry and Biology}},\
	}\href@noop {} {\bibfield  {journal} {\bibinfo  {journal} {phys. stat. sol.
				b}\ }\textbf {\bibinfo {volume} {217}},\ \bibinfo {pages} {41} (\bibinfo
		{year} {2000})}\BibitemShut {NoStop}%
	\bibitem [{\citenamefont {Koskinen}\ and\ \citenamefont
		{M{\"{a}}kinen}(2009)}]{koskinen_CMS_09}%
	\BibitemOpen
	\bibfield  {author} {\bibinfo {author} {\bibfnamefont {P.}~\bibnamefont
			{Koskinen}}\ and\ \bibinfo {author} {\bibfnamefont {V.}~\bibnamefont
			{M{\"{a}}kinen}},\ }\bibfield  {title} {\bibinfo {title} {{Density-functional
				tight-binding for beginners}},\ }\href@noop {} {\bibfield  {journal}
		{\bibinfo  {journal} {Comput. Mater. Sci.}\ }\textbf {\bibinfo {volume}
			{47}},\ \bibinfo {pages} {237} (\bibinfo {year} {2009})}\BibitemShut
	{NoStop}%
	\bibitem [{\citenamefont {Memarian}\ \emph {et~al.}(2015)\citenamefont
		{Memarian}, \citenamefont {Fereidoon},\ and\ \citenamefont {{Darvish
				Ganji}}}]{Memarian2015}%
	\BibitemOpen
	\bibfield  {author} {\bibinfo {author} {\bibfnamefont {F.}~\bibnamefont
			{Memarian}}, \bibinfo {author} {\bibfnamefont {A.}~\bibnamefont
			{Fereidoon}},\ and\ \bibinfo {author} {\bibfnamefont {M.}~\bibnamefont
			{{Darvish Ganji}}},\ }\bibfield  {title} {\bibinfo {title} {{Graphene Young's
				modulus: Molecular mechanics and DFT treatments}},\ }\href
	{https://doi.org/10.1016/j.spmi.2015.06.001} {\bibfield  {journal} {\bibinfo
			{journal} {Superlattices and Microstructures}\ }\textbf {\bibinfo {volume}
			{85}},\ \bibinfo {pages} {348} (\bibinfo {year} {2015})}\BibitemShut
	{NoStop}%
	\bibitem [{\citenamefont {Kudin}\ \emph {et~al.}(2001)\citenamefont {Kudin},
		\citenamefont {Scuseria},\ and\ \citenamefont {Yakobson}}]{kudin_PRB_01}%
	\BibitemOpen
	\bibfield  {author} {\bibinfo {author} {\bibfnamefont {K.~N.}\ \bibnamefont
			{Kudin}}, \bibinfo {author} {\bibfnamefont {G.~E.}\ \bibnamefont
			{Scuseria}},\ and\ \bibinfo {author} {\bibfnamefont {B.~I.}\ \bibnamefont
			{Yakobson}},\ }\bibfield  {title} {\bibinfo {title} {{C2F, BN, and C
				nanoshell elasticity from ab initio computations}},\ }\href
	{https://doi.org/10.1103/PhysRevB.64.235406} {\bibfield  {journal} {\bibinfo
			{journal} {Phys. Rev. B}\ }\textbf {\bibinfo {volume} {64}},\ \bibinfo
		{pages} {235406} (\bibinfo {year} {2001})}\BibitemShut {NoStop}%
	\bibitem [{\citenamefont {Koskinen}\ and\ \citenamefont
		{Kit}(2010{\natexlab{a}})}]{koskinen_PRB_10b}%
	\BibitemOpen
	\bibfield  {author} {\bibinfo {author} {\bibfnamefont {P.}~\bibnamefont
			{Koskinen}}\ and\ \bibinfo {author} {\bibfnamefont {O.~O.}\ \bibnamefont
			{Kit}},\ }\bibfield  {title} {\bibinfo {title} {{Approximate Modeling of
				Spherical Membranes}},\ }\href {https://doi.org/10.1103/PhysRevB.82.235420}
	{\bibfield  {journal} {\bibinfo  {journal} {Phys. Rev. B}\ }\textbf {\bibinfo
			{volume} {82}},\ \bibinfo {pages} {235420} (\bibinfo {year}
		{2010}{\natexlab{a}})}\BibitemShut {NoStop}%
	\bibitem [{\citenamefont {Frauenheim}\ \emph {et~al.}(2002)\citenamefont
		{Frauenheim}, \citenamefont {Seifert},\ and\ \citenamefont
		{Elstner}}]{Frauenheim2002}%
	\BibitemOpen
	\bibfield  {author} {\bibinfo {author} {\bibfnamefont {T.}~\bibnamefont
			{Frauenheim}}, \bibinfo {author} {\bibfnamefont {G.}~\bibnamefont
			{Seifert}},\ and\ \bibinfo {author} {\bibfnamefont {M.}~\bibnamefont
			{Elstner}},\ }\bibfield  {title} {\bibinfo {title} {{Atomistic simulations of
				complex materials: ground-state and excited-state properties}},\ }\href
	{http://iopscience.iop.org/0953-8984/14/11/313} {\bibfield  {journal}
		{\bibinfo  {journal} {J. Phys.: Condens. Matter}\ }\textbf {\bibinfo {volume}
			{14}},\ \bibinfo {pages} {3015} (\bibinfo {year} {2002})}\BibitemShut
	{NoStop}%
	\bibitem [{\citenamefont {Koskinen}(2010)}]{koskinen_PRB_10}%
	\BibitemOpen
	\bibfield  {author} {\bibinfo {author} {\bibfnamefont {P.}~\bibnamefont
			{Koskinen}},\ }\bibfield  {title} {\bibinfo {title} {{Electronic and optical
				properties of carbon nanotubes under pure bending}},\ }\href@noop {}
	{\bibfield  {journal} {\bibinfo  {journal} {Phys. Rev. B}\ }\textbf {\bibinfo
			{volume} {82}},\ \bibinfo {pages} {193409} (\bibinfo {year}
		{2010})}\BibitemShut {NoStop}%
	\bibitem [{\citenamefont {Koskinen}\ and\ \citenamefont
		{Kit}(2010{\natexlab{b}})}]{koskinen_PRL_10}%
	\BibitemOpen
	\bibfield  {author} {\bibinfo {author} {\bibfnamefont {P.}~\bibnamefont
			{Koskinen}}\ and\ \bibinfo {author} {\bibfnamefont {O.~O.}\ \bibnamefont
			{Kit}},\ }\bibfield  {title} {\bibinfo {title} {{Efficient approach for
				simulating distorted materials}},\ }\href@noop {} {\bibfield  {journal}
		{\bibinfo  {journal} {Phys. Rev. Lett.}\ }\textbf {\bibinfo {volume} {105}},\
		\bibinfo {pages} {106401} (\bibinfo {year} {2010}{\natexlab{b}})}\BibitemShut
	{NoStop}%
	\bibitem [{\citenamefont {Koskinen}(2012)}]{koskinen_PRB_12}%
	\BibitemOpen
	\bibfield  {author} {\bibinfo {author} {\bibfnamefont {P.}~\bibnamefont
			{Koskinen}},\ }\bibfield  {title} {\bibinfo {title} {{Graphene nanoribbons
				subject to gentle bends}},\ }\href
	{https://doi.org/10.1103/PhysRevB.85.205429} {\bibfield  {journal} {\bibinfo
			{journal} {Physical Review B}\ }\textbf {\bibinfo {volume} {85}},\ \bibinfo
		{pages} {205429} (\bibinfo {year} {2012})}\BibitemShut {NoStop}%
	\bibitem [{\citenamefont {Kit}\ \emph {et~al.}(2012)\citenamefont {Kit},
		\citenamefont {Tallinen}, \citenamefont {Mahadevan}, \citenamefont
		{Timonen},\ and\ \citenamefont {Koskinen}}]{kit_PRB_12}%
	\BibitemOpen
	\bibfield  {author} {\bibinfo {author} {\bibfnamefont {O.~O.}\ \bibnamefont
			{Kit}}, \bibinfo {author} {\bibfnamefont {T.}~\bibnamefont {Tallinen}},
		\bibinfo {author} {\bibfnamefont {L.}~\bibnamefont {Mahadevan}}, \bibinfo
		{author} {\bibfnamefont {J.}~\bibnamefont {Timonen}},\ and\ \bibinfo {author}
		{\bibfnamefont {P.}~\bibnamefont {Koskinen}},\ }\bibfield  {title} {\bibinfo
		{title} {{Twisting Graphene Nanoribbons into Carbon Nanotubes}},\ }\href
	{https://doi.org/10.1103/PhysRevB.85.085428} {\bibfield  {journal} {\bibinfo
			{journal} {Phys. Rev. B}\ }\textbf {\bibinfo {volume} {85}},\ \bibinfo
		{pages} {085428} (\bibinfo {year} {2012})}\BibitemShut {NoStop}%
	\bibitem [{\citenamefont {Ramasubramaniam}\ \emph {et~al.}(2012)\citenamefont
		{Ramasubramaniam}, \citenamefont {Koskinen}, \citenamefont {Kit},\ and\
		\citenamefont {Shenoy}}]{Ramasubramaniam2012}%
	\BibitemOpen
	\bibfield  {author} {\bibinfo {author} {\bibfnamefont {A.}~\bibnamefont
			{Ramasubramaniam}}, \bibinfo {author} {\bibfnamefont {P.}~\bibnamefont
			{Koskinen}}, \bibinfo {author} {\bibfnamefont {O.~O.}\ \bibnamefont {Kit}},\
		and\ \bibinfo {author} {\bibfnamefont {V.~B.}\ \bibnamefont {Shenoy}},\
	}\bibfield  {title} {\bibinfo {title} {{Edge-stress-induced spontaneous
				twisting of graphene nanoribbons}},\ }\href
	{https://doi.org/10.1063/1.3689814} {\bibfield  {journal} {\bibinfo
			{journal} {Journal of Applied Physics}\ }\textbf {\bibinfo {volume} {111}},\
		\bibinfo {pages} {054302} (\bibinfo {year} {2012})}\BibitemShut {NoStop}%
	\bibitem [{\citenamefont {Koskinen}(2011)}]{koskinen_APL_11}%
	\BibitemOpen
	\bibfield  {author} {\bibinfo {author} {\bibfnamefont {P.}~\bibnamefont
			{Koskinen}},\ }\bibfield  {title} {\bibinfo {title} {{Electromechanics of
				twisted graphene nanoribbons}},\ }\href@noop {} {\bibfield  {journal}
		{\bibinfo  {journal} {Appl. Phys. Lett.}\ }\textbf {\bibinfo {volume} {99}},\
		\bibinfo {pages} {013105} (\bibinfo {year} {2011})}\BibitemShut {NoStop}%
	\bibitem [{\citenamefont {Kit}\ \emph {et~al.}(2011)\citenamefont {Kit},
		\citenamefont {Pastewka},\ and\ \citenamefont {Koskinen}}]{kit_PRB_11}%
	\BibitemOpen
	\bibfield  {author} {\bibinfo {author} {\bibfnamefont {O.~O.}\ \bibnamefont
			{Kit}}, \bibinfo {author} {\bibfnamefont {L.}~\bibnamefont {Pastewka}},\ and\
		\bibinfo {author} {\bibfnamefont {P.}~\bibnamefont {Koskinen}},\ }\bibfield
	{title} {\bibinfo {title} {{Revised periodic boundary conditions:
				fundamentals, electrostatics and the tight-binding approximation}},\ }\href
	{https://doi.org/10.1103/PhysRevB.84.155431} {\bibfield  {journal} {\bibinfo
			{journal} {Phys. Rev. B}\ }\textbf {\bibinfo {volume} {84}},\ \bibinfo
		{pages} {155431} (\bibinfo {year} {2011})}\BibitemShut {NoStop}%
	\bibitem [{\citenamefont {Korhonen}\ and\ \citenamefont
		{Koskinen}(2014)}]{Korhonen2014a}%
	\BibitemOpen
	\bibfield  {author} {\bibinfo {author} {\bibfnamefont {T.}~\bibnamefont
			{Korhonen}}\ and\ \bibinfo {author} {\bibfnamefont {P.}~\bibnamefont
			{Koskinen}},\ }\bibfield  {title} {\bibinfo {title} {{Electromechanics of
				graphene spirals}},\ }\href {https://doi.org/10.1063/1.4904219} {\bibfield
		{journal} {\bibinfo  {journal} {AIP Advances}\ }\textbf {\bibinfo {volume}
			{4}},\ \bibinfo {pages} {127125} (\bibinfo {year} {2014})}\BibitemShut
	{NoStop}%
	\bibitem [{\citenamefont {Koskinen}(2016)}]{Koskinen2016}%
	\BibitemOpen
	\bibfield  {author} {\bibinfo {author} {\bibfnamefont {P.}~\bibnamefont
			{Koskinen}},\ }\bibfield  {title} {\bibinfo {title} {{Quantum Simulations of
				One-Dimensional Nanostructures under Arbitrary Deformations}},\ }\href
	{https://doi.org/10.1103/PhysRevApplied.6.034014} {\bibfield  {journal}
		{\bibinfo  {journal} {Physical Review Applied}\ }\textbf {\bibinfo {volume}
			{6}},\ \bibinfo {pages} {034014} (\bibinfo {year} {2016})}\BibitemShut
	{NoStop}%
	\bibitem [{\citenamefont {Bitzek}\ \emph {et~al.}(2006)\citenamefont {Bitzek},
		\citenamefont {Koskinen}, \citenamefont {G{\"{a}}hler}, \citenamefont
		{Moseler},\ and\ \citenamefont {Gumbsch}}]{bitzek_PRL_06}%
	\BibitemOpen
	\bibfield  {author} {\bibinfo {author} {\bibfnamefont {E.}~\bibnamefont
			{Bitzek}}, \bibinfo {author} {\bibfnamefont {P.}~\bibnamefont {Koskinen}},
		\bibinfo {author} {\bibfnamefont {F.}~\bibnamefont {G{\"{a}}hler}}, \bibinfo
		{author} {\bibfnamefont {M.}~\bibnamefont {Moseler}},\ and\ \bibinfo {author}
		{\bibfnamefont {P.}~\bibnamefont {Gumbsch}},\ }\bibfield  {title} {\bibinfo
		{title} {{Structural Relaxation Made Simple}},\ }\href
	{https://doi.org/10.1103/PhysRevLett.97.170201} {\bibfield  {journal}
		{\bibinfo  {journal} {Phys. Rev. Lett.}\ }\textbf {\bibinfo {volume} {97}},\
		\bibinfo {pages} {170201} (\bibinfo {year} {2006})}\BibitemShut {NoStop}%
	\bibitem [{\citenamefont {Koskinen}\ \emph {et~al.}(2008)\citenamefont
		{Koskinen}, \citenamefont {Malola},\ and\ \citenamefont
		{H{\"{a}}kkinen}}]{Koskinen2008}%
	\BibitemOpen
	\bibfield  {author} {\bibinfo {author} {\bibfnamefont {P.}~\bibnamefont
			{Koskinen}}, \bibinfo {author} {\bibfnamefont {S.}~\bibnamefont {Malola}},\
		and\ \bibinfo {author} {\bibfnamefont {H.}~\bibnamefont {H{\"{a}}kkinen}},\
	}\bibfield  {title} {\bibinfo {title} {{Self-Passivating Edge Reconstructions
				of Graphene}},\ }\href {https://doi.org/10.1103/PhysRevLett.101.115502}
	{\bibfield  {journal} {\bibinfo  {journal} {Phys. Rev. Lett.}\ }\textbf
		{\bibinfo {volume} {101}},\ \bibinfo {pages} {115502} (\bibinfo {year}
		{2008})}\BibitemShut {NoStop}%
	\bibitem [{\citenamefont {Koskinen}\ \emph {et~al.}(2009)\citenamefont
		{Koskinen}, \citenamefont {Malola},\ and\ \citenamefont
		{H{\"{a}}kkinen}}]{Koskinen2009a}%
	\BibitemOpen
	\bibfield  {author} {\bibinfo {author} {\bibfnamefont {P.}~\bibnamefont
			{Koskinen}}, \bibinfo {author} {\bibfnamefont {S.}~\bibnamefont {Malola}},\
		and\ \bibinfo {author} {\bibfnamefont {H.}~\bibnamefont {H{\"{a}}kkinen}},\
	}\bibfield  {title} {\bibinfo {title} {{Evidence for graphene edges beyond
				zigzag and armchair}},\ }\href {https://doi.org/10.1103/PhysRevB.80.073401}
	{\bibfield  {journal} {\bibinfo  {journal} {Physical Review B}\ }\textbf
		{\bibinfo {volume} {80}},\ \bibinfo {pages} {073401} (\bibinfo {year}
		{2009})}\BibitemShut {NoStop}%
	\bibitem [{\citenamefont {Seung}\ and\ \citenamefont
		{Nelson}(1988)}]{PhysRevA.38.1005}%
	\BibitemOpen
	\bibfield  {author} {\bibinfo {author} {\bibfnamefont {H.~S.}\ \bibnamefont
			{Seung}}\ and\ \bibinfo {author} {\bibfnamefont {D.~R.}\ \bibnamefont
			{Nelson}},\ }\bibfield  {title} {\bibinfo {title} {Defects in flexible
			membranes with crystalline order},\ }\href
	{https://doi.org/10.1103/PhysRevA.38.1005} {\bibfield  {journal} {\bibinfo
			{journal} {Phys. Rev. A}\ }\textbf {\bibinfo {volume} {38}},\ \bibinfo
		{pages} {1005} (\bibinfo {year} {1988})}\BibitemShut {NoStop}%
	\bibitem [{\citenamefont {Koskinen}\ \emph {et~al.}(2014)\citenamefont
		{Koskinen}, \citenamefont {Fampiou},\ and\ \citenamefont
		{Ramasubramaniam}}]{Koskinen2014a}%
	\BibitemOpen
	\bibfield  {author} {\bibinfo {author} {\bibfnamefont {P.}~\bibnamefont
			{Koskinen}}, \bibinfo {author} {\bibfnamefont {I.}~\bibnamefont {Fampiou}},\
		and\ \bibinfo {author} {\bibfnamefont {A.}~\bibnamefont {Ramasubramaniam}},\
	}\bibfield  {title} {\bibinfo {title} {Density-functional tight-binding
			simulations of curvature-controlled layer decoupling and band-gap tuning in
			bilayer mos2},\ }\href {https://doi.org/10.1103/PhysRevLett.112.186802}
	{\bibfield  {journal} {\bibinfo  {journal} {Physical Review Letters}\
		}\textbf {\bibinfo {volume} {112}},\ \bibinfo {pages} {186802} (\bibinfo
		{year} {2014})}\BibitemShut {NoStop}%
	\bibitem [{\citenamefont {Yu}\ \emph {et~al.}(2016)\citenamefont {Yu},
		\citenamefont {Ruzsinszky},\ and\ \citenamefont {Perdew}}]{Yu2016}%
	\BibitemOpen
	\bibfield  {author} {\bibinfo {author} {\bibfnamefont {L.}~\bibnamefont
			{Yu}}, \bibinfo {author} {\bibfnamefont {A.}~\bibnamefont {Ruzsinszky}},\
		and\ \bibinfo {author} {\bibfnamefont {J.~P.}\ \bibnamefont {Perdew}},\
	}\bibfield  {title} {\bibinfo {title} {{Bending Two-Dimensional Materials To
				Control Charge Localization and Fermi-Level Shift}},\ }\href
	{https://doi.org/10.1021/acs.nanolett.5b05303} {\bibfield  {journal}
		{\bibinfo  {journal} {Nano Letters}\ }\textbf {\bibinfo {volume} {16}},\
		\bibinfo {pages} {2444} (\bibinfo {year} {2016})}\BibitemShut {NoStop}%
	\bibitem [{\citenamefont {{Virtanen}}\ \emph {et~al.}(2020)\citenamefont
		{{Virtanen}}, \citenamefont {{Gommers}}, \citenamefont {{Oliphant}},
		\citenamefont {{Haberland}}, \citenamefont {{Reddy}}, \citenamefont
		{{Cournapeau}}, \citenamefont {{Burovski}}, \citenamefont {{Peterson}},
		\citenamefont {{Weckesser}}, \citenamefont {{Bright}}, \citenamefont {{van
				der Walt}}, \citenamefont {{Brett}}, \citenamefont {{Wilson}}, \citenamefont
		{{Jarrod Millman}}, \citenamefont {{Mayorov}}, \citenamefont {{Nelson}},
		\citenamefont {{Jones}}, \citenamefont {{Kern}}, \citenamefont {{Larson}},
		\citenamefont {{Carey}}, \citenamefont {{Polat}}, \citenamefont {{Feng}},
		\citenamefont {{Moore}}, \citenamefont {{Vand erPlas}}, \citenamefont
		{{Laxalde}}, \citenamefont {{Perktold}}, \citenamefont {{Cimrman}},
		\citenamefont {{Henriksen}}, \citenamefont {{Quintero}}, \citenamefont
		{{Harris}}, \citenamefont {{Archibald}}, \citenamefont {{Ribeiro}},
		\citenamefont {{Pedregosa}}, \citenamefont {{van Mulbregt}},\ and\
		\citenamefont {{Contributors}}}]{2020SciPy-NMeth}%
	\BibitemOpen
	\bibfield  {author} {\bibinfo {author} {\bibfnamefont {P.}~\bibnamefont
			{{Virtanen}}}, \bibinfo {author} {\bibfnamefont {R.}~\bibnamefont
			{{Gommers}}}, \bibinfo {author} {\bibfnamefont {T.~E.}\ \bibnamefont
			{{Oliphant}}}, \bibinfo {author} {\bibfnamefont {M.}~\bibnamefont
			{{Haberland}}}, \bibinfo {author} {\bibfnamefont {T.}~\bibnamefont
			{{Reddy}}}, \bibinfo {author} {\bibfnamefont {D.}~\bibnamefont
			{{Cournapeau}}}, \bibinfo {author} {\bibfnamefont {E.}~\bibnamefont
			{{Burovski}}}, \bibinfo {author} {\bibfnamefont {P.}~\bibnamefont
			{{Peterson}}}, \bibinfo {author} {\bibfnamefont {W.}~\bibnamefont
			{{Weckesser}}}, \bibinfo {author} {\bibfnamefont {J.}~\bibnamefont
			{{Bright}}}, \bibinfo {author} {\bibfnamefont {S.~J.}\ \bibnamefont {{van der
					Walt}}}, \bibinfo {author} {\bibfnamefont {M.}~\bibnamefont {{Brett}}},
		\bibinfo {author} {\bibfnamefont {J.}~\bibnamefont {{Wilson}}}, \bibinfo
		{author} {\bibfnamefont {K.}~\bibnamefont {{Jarrod Millman}}}, \bibinfo
		{author} {\bibfnamefont {N.}~\bibnamefont {{Mayorov}}}, \bibinfo {author}
		{\bibfnamefont {A.~R.~J.}\ \bibnamefont {{Nelson}}}, \bibinfo {author}
		{\bibfnamefont {E.}~\bibnamefont {{Jones}}}, \bibinfo {author} {\bibfnamefont
			{R.}~\bibnamefont {{Kern}}}, \bibinfo {author} {\bibfnamefont
			{E.}~\bibnamefont {{Larson}}}, \bibinfo {author} {\bibfnamefont
			{C.}~\bibnamefont {{Carey}}}, \bibinfo {author} {\bibfnamefont
			{{\.I}.}~\bibnamefont {{Polat}}}, \bibinfo {author} {\bibfnamefont
			{Y.}~\bibnamefont {{Feng}}}, \bibinfo {author} {\bibfnamefont {E.~W.}\
			\bibnamefont {{Moore}}}, \bibinfo {author} {\bibfnamefont {J.}~\bibnamefont
			{{Vand erPlas}}}, \bibinfo {author} {\bibfnamefont {D.}~\bibnamefont
			{{Laxalde}}}, \bibinfo {author} {\bibfnamefont {J.}~\bibnamefont
			{{Perktold}}}, \bibinfo {author} {\bibfnamefont {R.}~\bibnamefont
			{{Cimrman}}}, \bibinfo {author} {\bibfnamefont {I.}~\bibnamefont
			{{Henriksen}}}, \bibinfo {author} {\bibfnamefont {E.~A.}\ \bibnamefont
			{{Quintero}}}, \bibinfo {author} {\bibfnamefont {C.~R.}\ \bibnamefont
			{{Harris}}}, \bibinfo {author} {\bibfnamefont {A.~M.}\ \bibnamefont
			{{Archibald}}}, \bibinfo {author} {\bibfnamefont {A.~H.}\ \bibnamefont
			{{Ribeiro}}}, \bibinfo {author} {\bibfnamefont {F.}~\bibnamefont
			{{Pedregosa}}}, \bibinfo {author} {\bibfnamefont {P.}~\bibnamefont {{van
					Mulbregt}}},\ and\ \bibinfo {author} {\bibfnamefont {S.~.~.}\ \bibnamefont
			{{Contributors}}},\ }\bibfield  {title} {\bibinfo {title} {{SciPy 1.0:
				Fundamental Algorithms for Scientific Computing in Python}},\ }\href
	{https://doi.org/https://doi.org/10.1038/s41592-019-0686-2} {\bibfield
		{journal} {\bibinfo  {journal} {Nature Methods}\ }\textbf {\bibinfo {volume}
			{17}},\ \bibinfo {pages} {261} (\bibinfo {year} {2020})}\BibitemShut
	{NoStop}%
	\bibitem [{\citenamefont {Shenoy}\ \emph {et~al.}(2008)\citenamefont {Shenoy},
		\citenamefont {Reddy}, \citenamefont {Ramasubramaniam},\ and\ \citenamefont
		{Zhang}}]{Shenoy2008}%
	\BibitemOpen
	\bibfield  {author} {\bibinfo {author} {\bibfnamefont {V.~B.}\ \bibnamefont
			{Shenoy}}, \bibinfo {author} {\bibfnamefont {C.~D.}\ \bibnamefont {Reddy}},
		\bibinfo {author} {\bibfnamefont {A.}~\bibnamefont {Ramasubramaniam}},\ and\
		\bibinfo {author} {\bibfnamefont {Y.~W.}\ \bibnamefont {Zhang}},\ }\bibfield
	{title} {\bibinfo {title} {{Edge-Stress-Induced Warping of Graphene Sheets
				and Nanoribbons}},\ }\href {https://doi.org/10.1103/PhysRevLett.101.245501}
	{\bibfield  {journal} {\bibinfo  {journal} {Phys. Rev. Lett.}\ }\textbf
		{\bibinfo {volume} {101}},\ \bibinfo {pages} {245501} (\bibinfo {year}
		{2008})}\BibitemShut {NoStop}%
	\bibitem [{\citenamefont {Shenoy}\ \emph {et~al.}(2010)\citenamefont {Shenoy},
		\citenamefont {Reddy},\ and\ \citenamefont {Zhang}}]{Shenoy2010}%
	\BibitemOpen
	\bibfield  {author} {\bibinfo {author} {\bibfnamefont {V.~B.}\ \bibnamefont
			{Shenoy}}, \bibinfo {author} {\bibfnamefont {C.~D.}\ \bibnamefont {Reddy}},\
		and\ \bibinfo {author} {\bibfnamefont {Y.-W.}\ \bibnamefont {Zhang}},\
	}\bibfield  {title} {\bibinfo {title} {{Spontaneous curling of graphene
				sheets with reconstructed edges.}},\ }\href
	{https://doi.org/10.1021/nn100842k} {\bibfield  {journal} {\bibinfo
			{journal} {ACS Nano}\ }\textbf {\bibinfo {volume} {4}},\ \bibinfo {pages}
		{4840} (\bibinfo {year} {2010})}\BibitemShut {NoStop}%
	\bibitem [{\citenamefont {Balandin}(2011)}]{balandin_nmat_11}%
	\BibitemOpen
	\bibfield  {author} {\bibinfo {author} {\bibfnamefont {A.~A.}\ \bibnamefont
			{Balandin}},\ }\bibfield  {title} {\bibinfo {title} {{Thermal properties of
				graphene and nanostructured carbon materials}},\ }\href@noop {} {\bibfield
		{journal} {\bibinfo  {journal} {Nature Mat.}\ }\textbf {\bibinfo {volume}
			{10}},\ \bibinfo {pages} {569} (\bibinfo {year} {2011})}\BibitemShut
	{NoStop}%
\end{thebibliography}

%

\end{document}